\RequirePackage{ifpdf}
\ifpdf 
\documentclass[pdftex]{sigma}
\else
\documentclass{sigma}
\fi

\numberwithin{equation}{section}

\usepackage{epsf,bm}

\newcommand{\modS}{ {\cal S}}
\newcommand{\modT}{ {\cal T}}

\begin{document}
\allowdisplaybreaks

\renewcommand{\PaperNumber}{010}

\FirstPageHeading

\renewcommand{\thefootnote}{$\star$}

\ShortArticleName{Duality in Supersymmetric Yang--Mills and the Quantum Hall Ef\/fect}

\ArticleName{Modular Symmetry and Fractional Charges \\
in $\boldsymbol{N=2}$ Supersymmetric Yang--Mills\\ and the Quantum 
Hall Ef\/fect\footnote{This paper is a contribution to the Proceedings of
the O'Raifeartaigh Symposium on Non-Perturbative and Symmetry
Methods in Field Theory
 (June 22--24, 2006, Budapest, Hungary).
The full collection is available at
\href{http://www.emis.de/journals/SIGMA/LOR2006.html}{http://www.emis.de/journals/SIGMA/LOR2006.html}}}

\Author{Brian P. DOLAN~$^{\dag^1\dag^2}$}
\AuthorNameForHeading{B.P. Dolan}

\Address{$^{\dag^1}$~Department of Mathematical Physics, National University of Ireland, Maynooth, Ireland}

\Address{$^{\dag^2}$~School of Theoretical Physics,
Dublin Institute for Advanced Studies,\\
$\phantom{^{\dag^2}}$~10, Burlington Rd., Dublin, Ireland} 

\EmailDD{\href{mailto:bdolan@thphys.nuim.ie}{bdolan@thphys.nuim.ie}}
\URLaddressDD{\url{http://www.thphys.nuim.ie/staff/bdolan/}}

\ArticleDates{Received September 29, 2006, in f\/inal form December
22, 2006; Published online January 10, 2007}

\Abstract{The parallel r\^oles of modular symmetry in ${\cal N}=2$ supersymmetric Yang--Mills
and in the quantum Hall ef\/fect are reviewed. 
In supersymmetric Yang--Mills theories modular symmetry emerges as 
a version of Dirac's electric -- magnetic duality. It has
signif\/icant consequences for the vacuum structure of these theories, leading to
a fractal vacuum which has an inf\/inite hierarchy of related phases.
In the case of ${\cal N}=2$ supersymmetric Yang--Mills in $3+1$ dimensions, scaling
functions can be def\/ined which are modular forms of a~subgroup of the full modular group and
which interpolate between vacua. 
Infra-red f\/ixed points at strong coupling correspond to $\theta$-vacua with $\theta$ a
rational number that, in the case of pure SUSY Yang--Mills, has odd denominator.
There is a mass gap for electrically charged particles which can carry 
fractional electric charge.
A similar structure applies to the $2+1$ dimensional quantum Hall ef\/fect where the hierarchy of Hall
plateaux can be understood in terms of an action of the modular group and the stability
of Hall plateaux is due to the fact that odd denominator
Hall conductivities are attractive infra-red f\/ixed points.
There is a mass gap for electrically
charged excitations which, in the case of the fractional quantum Hall ef\/fect, carry
fractional electric charge.}


\Classification{11F11; 81R05; 81T60; 81V70}

\section{Introduction}

The r\^ole of the modular group as a duality symmetry in physics has gained increasing
prominence in recent years, not only through string theory considerations and 
in supersymmetric gauge theories but also in 2+1 dimensional $U(1)$ gauge theories.
In this last case there is considerable contact with experiment via the quantum Hall ef\/fect, where modular symmetry relates the integer and fractional quantum Hall ef\/fects.
This article is a review of the current understanding of modular symmetry in
the quantum Hall ef\/fect and the remarkable similarities
with 3+1 dimensional ${\cal N}=2$ supersymmetric Yang--Mills theory.
It is an extended version of the mini-review \cite{minireview}.

Mathematically the connection between ${\cal N}=2$ supersymmetric $SU(2)$ Yang--Mills
and the quantum Hall ef\/fect lies in the constraints that modular symmetry places
on the scaling f\/low of the two theories,
as will be elucidated below.  The quantum Hall ef\/fect is of course a real physical
system with imperfections and impurities and as such modular symmetry can only
be approximate in the phenomena, 
nevertheless there is strong experimental support for the relevance of modular symmetry
in the experimental data. 

The signif\/icance of the modular group, as a generalisation of electric-magnetic duality,
was emphasised by Shapere and Wilczek \cite{WilczekShapere}, and its full power was realised
by Seiberg and Witten in supersymmetric Yang--Mills,
motivated in part by the Montonen and Olive conjecture \cite{MO}.
In the case of ${\cal N}=2$ SUSY Yang--Mills 
Seiberg and Witten showed, in their seminal papers in 1994~\cite{SW,SW2}, 
that a remnant of the full modular group survives in the low energy physics.

The earliest appearance of modular symmetry in the condensed matter literature was
in the work of Cardy and Rabinovici, \cite{CardyRab,Cardy} where a coupled clock model 
was analysed,
interestingly with a view to gaining insight into quantum chromodynamics.  
A suggested link between phase diagrams
for one-dimensional clock models and the quantum Hall ef\/fect was made in \cite{Asorey}.  
An action of the modular group leading to a fractal structure was
found in applying dissipative quantum mechanics to the Hofstadter model, \cite{Callan1,Callan2}
and fractal structures have also emerged in other models of the quantum Hall ef\/fect \cite{Welly,Welly+}.
The f\/irst mention of the modular group in relation to the quantum Hall ef\/fect appears
to have been by Wilczek and Shapere in \cite{WilczekShapere}, 
but these authors focused on a particular subgroup 
which has f\/ixed points that are not observed in the experimental data on 
the quantum Hall ef\/fect.
A more detailed analysis was undertaken by L\"utken and Ross \cite{LutkenRoss1,LutkenRoss2} and the correct
subgroup was f\/inally identif\/ied unambiguously 
in~\cite{LutkenRoss2,Lutken,Lutken+}, at least for spin-split quantum Hall samples.
Almost at the same time as L\"utken and Ross' paper Kivelson, Lee and Zhang derived
their ``Law of Corresponding States'' \cite{KLZ}, 
relating dif\/ferent quantum Hall plateaux
in spin-split samples.
Although they did not mention modular symmetry in their paper, 
their map is in fact the group $\Gamma_0(2)$ described below and discussed 
in~\cite{LutkenRoss2} and \cite{Lutken,Lutken+}. 

In Section~2 duality in electromagnetism is reviewed and the Dirac--Schwinger--Zwanziger
quantisation condition and the Witten ef\/fect are described. 
Section 3 introduces the
subgroups of the modular group that are relevant to ${\cal N}=2$ SUSY $SU(2)$
Yang--Mills and scaling functions, modular forms that are regular at all 
the singular points in the moduli space of vacua, are discussed.
In Section~4 the relevance of the modular group to the quantum Hall ef\/fect is described and
Kivelson, Lee and Zhang's derivation of the Law of Corresponding States, for 
spin-split quantum Hall samples, and its relation to the modular group, is explained.
Modular symmetries of spin-degenerate samples are also reviewed and predictions
for hierarchical structures in bosonic systems, based on a dif\/ferent subgroup of the
modular group to that of the QHE, is also explained.
Finally Section~5 contains a summary and conclusions.

\renewcommand{\thefootnote}{\arabic{footnote}}
\setcounter{footnote}{0}

\section{Duality in electromagnetism}

Maxwell's equations in the absence of sources 
\begin{gather*}
\nabla \times {\bf E} + \frac{\partial {\bf B}}{\partial t} =0,\qquad \nabla\cdot {\bf E}=0, \nonumber \\
\nabla \times {\bf B} - \frac{\partial {\bf E}}{\partial t} =0,\qquad \nabla\cdot {\bf B}=0 
\end{gather*}
(using units in which $\epsilon_0=\mu_0=c=1$) are not only symmetric under the conformal group
in 3+1 dimensions but also under
the interchange of electric f\/ield $\bf E$ and the magnetic f\/ield $\bf B$,
more specif\/ically Maxwell's equations are symmetric under the map 
\begin{gather}\label{Duality}
{\bf E}\rightarrow {\bf B} \qquad\hbox{and}\qquad {\bf B}\rightarrow -{\bf E}.
\end{gather}
This symmetry is known as {\it duality}, for any f\/ield conf\/iguration $({\bf E},{\bf B})$ there is
a dual conf\/iguration $({\bf B},-{\bf E})$
(a very good introduction to these ideas is~\cite{Jose}).
Duality is a useful symmetry, e.g.\ in rectangular wave-guide problems,
using this symmetry one can immediately construct
transverse magnetic modes once transverse electric modes are known.

In fact there is a larger continuous symmetry which is perhaps most easily seen by def\/ining the
complex f\/ield
\begin{gather*}
{\bf F}={\bf B}+i{\bf E}
\end{gather*}
and writing the source free Maxwell's equations as
\begin{gather*} 
\nabla \times {\bf F}+i\frac{\partial {\bf F}}{\partial t}=0,\qquad \nabla \cdot {\bf F}=0.
\end{gather*}
Then the map ${\bf F}\rightarrow e^{i\phi}{\bf F}$, with $\phi$ a constant phase,
is also a symmetry of the vacuum Maxwell equations sending
\begin{gather*}
{\bf B}\rightarrow \cos\phi\,{\bf B}-\sin\phi\,{\bf E}\qquad\hbox{and}\qquad
{\bf E}\rightarrow \cos\phi\,{\bf E}+\sin\phi\,{\bf B}, 
\end{gather*}
the particular case of $\phi=\pi/2$ giving (\ref{Duality}) above.

When electric sources, i.e.\ a current $J^\mu$, are included this is no longer a symmetry but,  
in a~seminal paper \cite{DQC}, Dirac showed that 
a vestige of (\ref{Duality}) remains provided magnetic monopoles are introduced,
or more generally magnetic currents, ${\widetilde J}^\mu$.
By quantising a charged particle, with electric charge $Q$, in a background magnetic monopole
f\/ield, generated by a monopole with charge $M$, Dirac showed that single-valuedness of the
wave-function requires that $QM$ must be an integral multiple of Planck's constant, or
\begin{gather} \label{DQC}
QM=2\pi \hbar n\end{gather}
where $n$ is an integer -- the famous Dirac quantisation condition which
is intimately connected with topology and the theory of f\/ibre bundles~\cite{WuYang}.

A quick way of deriving (\ref{DQC}) is to consider the orbital angular momentum of 
particle of mass ${\tt m}$ and charge $Q$ in the presence of a magnetic monopole $M$ generating
the magnetic f\/ield
\begin{gather*} 
{\bf B}= \frac{M}{4\pi} \frac{\bf r}{r^3}.
\end{gather*}
The form of the Lorentz force, ${\tt m}\ddot{\bf r} = Q(\dot{\bf r}\times {\bf B})$,
implies that orbital angular momentum, 
${\bf L} = {\tt m}({\bf r}\times \dot{\bf r})$, 
is not conserved
\begin{gather*} \dot {\bf L}={\tt m}({\bf r}\times \ddot{\bf r})={\bf r}\times 
\bigl(Q\bigl(\dot{\bf r}\times{\bf B}\bigr)\bigr)
= \frac{QM}{4\pi r^3} {\bf r} \times \bigl(\dot{\bf r}\times{\bf r}\bigr)=\frac{d}{dt}
\left(\frac{QM}{4\pi}\frac{\bf r}{r}\right).
\end{gather*}
However 
\begin{gather*}
{\bf J}:={\bf L}-\left(\frac{QM}{ 4\pi}\frac{\bf r}{r}\right),
\end{gather*}
is conserved and so we def\/ine this to be the total angular momentum of the particle plus the f\/ield.
Indeed 
\begin{gather*} {\bf J}^{\rm em}=\int {\bf r}'\times \bigl({\bf E}({\bf r}'
-{\bf r})\times {\bf B}({\bf r}')\bigr) d^3 r'=
-\left(\frac{QM}{4\pi}\frac{\bf r}{r}\right)
\end{gather*}
is the angular momentum of the electromagnetic f\/ield 
generated by a magnetic monopole $M$
separated from an electric charge $Q$ by ${\bf r}$.
Assuming that ${\bf J}^{\rm em}$ is quantised in the usual way the result of a measurement should 
always yield
\begin{gather*}
J^{\rm em}=\frac{1}{2}n\hbar,
\end{gather*}
where $n$ is an integer, leading to (\ref{DQC}).

The Dirac quantisation condition has the fascinating consequence that,
if a single magnetic monopole $M$ exists anywhere in the universe then
electric charge
\begin{gather*} 
Q=\frac{2\pi \hbar n}{M}
\end{gather*}
must be quantised as a multiple of $\frac{2\pi \hbar}{M}$.  Conversely if $Q=e$ is 
a fundamental unit of electric charge
then there is a unit of magnetic charge, namely
\begin{gather} \label{Munit} 
m=\frac{2\pi\hbar}{e},
\end{gather}
and the allowed magnetic charges are 
integral multiples of $m$.

Since magnetic monopoles have never been observed, if they exist at all, they must be very heavy,
much heavier than an electron\footnote{For a review of the current status of magnetic monopoles
see \cite{Milton}.}, and so duality is not a manifest symmetry of the physics~-- 
it is at best a map between two descriptions of the same theory.
In perturbation theory, since 
$\alpha_e=\frac{e^2}{4\pi\hbar}\approx\frac{1}{137}$ is so small, the `magnetic' f\/ine
structure constant
\begin{gather}\label{alphaem}
\alpha_m=\frac{m^2}{4\pi\hbar}=\frac{1}{4}\frac{1}{\alpha_e}\approx 34\end{gather}
is very large and magnetic monopoles have very strong coupling to the electromagnetic
f\/ield.  One could not construct a perturbation theory of magnetic monopoles, but one could analyse
a~theory of magnetic monopoles by f\/irst performing perturbative calculations of electric
charges and then using duality to map the results to the strongly coupled magnetic charges.
Duality is thus potentially a very useful tool in quantum f\/ield theory since it provides
a mathematical tool for studying strongly interacting theories.

There is a generalisation of the Dirac quantisation condition for particles that carry both an electric
and a magnetic charge at the same time, called {\it dyons}.  If a dyon with electric
charge~$Q$ and magnetic charge $M$ orbits a second dyon with charges $Q'$ and $M'$, then Schwinger 
and Zwanziger showed that \cite{Schwinger,Zwanziger}
\begin{gather*} 
QM'-Q'M=2\pi n \hbar.
\end{gather*}
A dimensionless version of the Dirac--Schwinger--Zwanziger quantisation condition
can be obtained by writing 
\begin{gather*} 
Q=n_e e, \qquad Q'=n_e' e, \qquad M=n_m m, \qquad \hbox{and}\qquad M'=n_m' m 
\end{gather*}
giving
\begin{gather}\label{SZ2}
n_e n_m'-n_e' n_m=n
\end{gather}  
with $n_e$, $n_e'$, $n_p$, $n_p'$ and $n$ integers.

The simple ${\bf Z}_2$ duality map (\ref{alphaem})
\begin{gather} \label{Zed2}
2\alpha_e\quad\rightarrow\quad 2\alpha_m =\frac{1}{2\alpha_e}\end{gather}
can be extended to a much richer structure \cite{WilczekShapere,WittenModular,WittenDuality,Verlinde} 
involving an inf\/inite discrete group, the modular group
$Sl(2,{\bf Z})/{{\bf Z}_2}\cong \Gamma(1)$, by including a topological term in the 4-dimensional
action,
\begin{gather} 
S =\int \left(-\frac{1}{4e^2}F_{\mu\nu}F^{\mu\nu} 
+\frac{\theta}{32 \pi^2} \epsilon^{\mu\nu\rho\sigma}F_{\mu\nu}F_{\rho\sigma} \right)d^4 x \nonumber \\
\phantom{S}=\int \left(-\frac{1}{2e^2}F\wedge *F +\frac{\theta}{ 8\pi^2} F\wedge F \right),\label{EMAction}
\end{gather}
with $F=\frac{1}{2}F_{\mu\nu}dx^\mu\wedge dx^\nu$ in dif\/ferential form notation.
Def\/ine the complex parameter
\begin{gather*} 
\tau = \frac{\theta}{2\pi} + \frac{2\pi i}{e^2}, 
\end{gather*}
using units with $\hbar =1$,\footnote{$\tau=\frac{\theta}{2\pi} + \frac{2\pi i\hbar }{e^2}$
when $\hbar$ is included.} and the combinations
\begin{gather*} 
F_\pm=\frac{1}{2}(F\pm i*F) \quad\Rightarrow \quad *F_\pm =\mp i F_\pm,
\end{gather*}
then the action can be written as
\begin{gather*} 
S=\frac{i}{4\pi}\int\left\{ \tau(F_+\wedge *F_+) - \overline\tau (F_-\wedge *F_-)\right\}. 
\end{gather*}
In terms of $\tau$ (\ref{Zed2}) generalises to a map that we shall denote by ${\cal S}$
\begin{gather*}
{\cal S}:\ \tau\quad\rightarrow \quad -\frac{1}{\tau},
\end{gather*}
which reduces to (\ref{Zed2}) when $\theta=0$.
Notice that the imaginary part of $\tau$ satisf\/ies $\Im \tau>0$ (for stability $e^2>0$)
and so $\tau$ parameterises the upper-half complex plane.

While it is well known that the topological term in (\ref{EMAction}) has interesting physical
consequences in non-Abelian gauge theories, leading to the $\theta$-vacua in 
QCD \cite{Poly,BPST,tHoofttheta,CDG,JR},
it is not usually considered in an Abelian theory.  
But it can have non-trivial consequences for the
partition function even for Abelian theories, for example if space-time is a 4-torus 
$T^4=T^2\times {\widetilde T}^2$
then the second term in (\ref{EMAction}) can be non-zero.  Consider 
for example a f\/ield conf\/iguration in which the 2-form $F$ is
the direct sum of a monopole with f\/irst Chern class $p_1$ on one 2-torus $T^2$ and 
a~monopole with f\/irst Chern class $p_2$ on the second ${\widetilde T}^2$, i.e.
\begin{gather*} F=F^{(1)} + F^{(2)}\end{gather*} 
with 
\begin{gather*} \frac{1}{2\pi}\int_{T^2}F^{(1)}=p_1
\end{gather*} 
on the f\/irst torus and
\begin{gather*} \frac{1}{2\pi}\int_{{\widetilde T}^2}F^{(2)}=p_2
\end{gather*} 
on the second.  Then
\begin{gather*} 
\frac{1}{8\pi^2}\int_{T^4} F \wedge F = \frac{1}{4\pi^2}\int_{T^2\times {\widetilde T}^2} F^{(1)}\wedge F^{(2)}
= \left(\frac{1}{2\pi}\int_{T^2} F^{(1)}\right)\left(\frac{1}{2\pi}\int_{{\widetilde T}^2} F^{(2)}\right)
 = p_1 p_2 
 \end{gather*}
is an integer and $e^{iS}$, and hence the partition function, 
is invariant under $\theta \rightarrow \theta + 2\pi$.  We thus def\/ine a second operation ${\cal T}$ on $\tau$
\begin{gather*} {\cal T}:\tau\quad\rightarrow \quad \tau+1.
\end{gather*}

Both $\modS$ and $\modT$ preserve the property that $\Im\tau >0$ and they generate the
modular group acting on the upper-half complex plane.  A general element of the
modular group can be represented as a $2\times 2$ matrix of integers with unit determinant,
\begin{gather} \label{abcd} \gamma=\left(
\begin{array}{cc}
a & b \\
 c & d 
\end{array}\right) \end{gather}
with $ad-bc=1$, and
\begin{gather} \label{modulartrans}
\gamma(\tau) =\frac{a\tau + b}{c\tau +d}.\end{gather}
Clearly $-\gamma$ has the same ef\/fect on $\tau$ as $\gamma$ does so the modular group is
$\Gamma(1)\cong Sl(2,{\bf Z})/{{\bf Z}_2}$.  In Euclidean space a consequence of
this is that the partition function is related to 
a modular form and depends on topological invariants, the Euler characteristic 
and the Hirzebruch signature, in a well-def\/ined manner, \cite{WittenDuality,Verlinde}.

Magnetic monopoles {\it \'a la} Dirac are singular at the origin but, remarkably, it is possible
to f\/ind classical solutions of coupled $SO(3)$ Yang--Mills--Higgs systems which are 
f\/inite and smooth
at the origin and reduce to monopole conf\/igurations, with $M=-\frac{4\pi\hbar}{e}$, 
at large $r$, 't Hooft--Polyakov monopoles \cite{tHooftMono,PolyakovMono}.  
The Higgs f\/ield $\Phi$, which is in the adjoint 
representation
of $SO(3)$, acquires a non-zero vacuum expectation value away from the origin,
${\Phi\cdot \Phi}={\bm a}^2$, which breaks
the gauge symmetry down to $U(1)$.
An interesting consequence of the inclusion of the $\theta$-term in the \hbox{'t~Hooft}--Polyakov action
was pointed out by Witten in \cite{WittenModular}.  
A necessary
condition for gauge invariance of the partition function is
\begin{gather*} 
\exp\left\{2\pi i\left(\frac{Q}{e} -\frac{\theta e}{8\pi^2}\frac{M}{\hbar}\right)\right\}=1
\qquad\Rightarrow\qquad
\frac{Q}{e} -\frac{\theta e}{8\pi^2}\frac{M}{\hbar}=n_e
\quad\hbox{with}\quad n_e\in{\bf Z}.
\end{gather*}
Allowing $M$ to be an integer multiple of the
't Hooft--Polyakov monopole charge
\begin{gather} \label{Mmn} M=n_m\left(\frac{4\pi\hbar}{e}\right),\end{gather}
(which dif\/fers from (\ref{Munit}) by a factor of 2 because the gauge group is $SO(3)$ and there
are no spinor representations allowed) 
we see that
\begin{gather} \label{WittenEffect}
\frac{Q}{e}=n_e + n_m\frac{\theta}{2\pi}\end{gather}
with $n_e$ and $n_m$ integers, the electric and magnetic quantum numbers
(for the 't Hooft--Polyakov monopole $n_e=0$ and $n_m=-1$).
This reduces to $Q=n_e e$ when $\theta=0$ as before.
If $\frac{\theta}{2\pi}=\frac{q}{p}$ is rational, with $q$ and $p$ mutually prime integers, 
then there exists the possibility of a magnetically charged particle 
with zero electric charge, when $(n_m,n_e)=(-p,q)$.  Furthermore, since~$q$ and~$p$ are mutually
prime, it is an elementary result of number theory that there also exist integers $(n_m,n_e)$
such that 
\begin{gather*} 
n_e p + n_m q =1\qquad\hbox{and hence}\qquad Q=\frac{e}{p},
\end{gather*}
i.e.\ the Witten ef\/fect allows for fractionally charge particles.
We shall see in Section 4 that the fractionally charged particles observed in the
quantum Hall ef\/fect are a 2+1-dimensional analogue of this.

The masses of dyons in this  model are related to the vacuum expectation value
of the Higgs f\/ield and the electric and magnetic quantum numbers,
in particular the masses ${\cal M}$ are bounded below by the Bogomol'nyi bound \cite{Jose,Bog}
\begin{gather} \label{BogBound} 
{\cal M}^2 \ge  \frac{{\bm a}^2}{e^2} (Q^2 + M^2).
\end{gather}

\section[Duality in ${\cal N}=2$, $SU(2)$ SUSY Yang-Mills]{Duality 
in $\boldsymbol{{\cal N}=2}$, $\boldsymbol{SU(2)}$ SUSY Yang--Mills}

So far the discussion has been semi-classical:
Dirac considered a quantised particle moving in
a f\/ixed classical background f\/ield and modular transformations are unlikely to be
useful in the full theory of QED when coupled electromagnetic and matter
f\/ields are quantised.
Nevertheless Seiberg and Witten showed in 1994 \cite{SW} that
in fully quantised 
${\cal N}=2$ supersymmetric Yang--Mills theory, where the non-Abelian gauge group is broken to $U(1)$,
modular transformations of the complex coupling, or at least a subgroup thereof, are manifest in the vacuum structure.
In essence supersymmetry is such a powerful constraint
that, in the low-energy/long-wavelength limit of the theory, a remnant of the semi-classical
modular action survives in the full quantum theory.

Consider $SU(2)$ Yang--Mills in 4-dimensional Minkowski space 
with global ${\cal N}=2$ supersymmetry and, in the simplest case,
no matter f\/ields\footnote{There is by now a number of good
reviews of ${\cal N}=2$ SUSY, see for example \cite{AlvarezG,Peskin}.}.  
The f\/ield content is then the $SU(2)$ gauge potential, $ A_{\mu}$; two Weyl
spinors (gauginos) in a doublet of ${\cal N}=2$ supersymmetry, 
both transforming under the adjoint representation of $SU(2)$, and a
single complex scalar $\phi$, again in the adjoint representation.  The 
bosonic part of the action is
\begin{gather*} 
S =
\int dx^4 \left\{-\frac{1}{2 g^2} \,{\rm tr}(F_{\mu\nu}F^{\mu\nu})+
\frac{\theta}{32 \pi^2} \varepsilon^{\mu\nu\rho\sigma}\,{\rm tr}(F_{\mu\nu}F_{\rho\sigma})\right. \nonumber \\
\left.\phantom{S=}{} +\frac{1}{g^2}
\,{\rm tr}\left(  \left (D_\mu\phi\right)^\dagger D^\mu\phi -\frac{1}{2}[\phi^\dagger,\phi]^2\right)\right\}.
\end{gather*} 
The fermionic terms (not exhibited explicitly here) 
are dictated by supersymmetry and involve Yukawa interactions with the scalar f\/ield
but it is crucial that supersymmetry relates all the couplings and the only free parameters are the gauge
coupling $g$ and the vacuum angle $\theta$, all other couplings are in fact determined by $g$.
Just as in electrodynamics these parameters can be combined into a single complex
parameter which, in the conventions of \cite{SW}, is
\begin{gather} \label{taudef2} \tau=\frac{\theta}{2\pi} +i\frac{4\pi}{g^2}.\end{gather}
(Note that the imaginary part of $\tau$ is def\/ined here to be twice that in Section~2 -- this is because 
the 't Hooft--Polyakov monopole
is central to the understanding of $N=2$ SUSY Yang--Mills and it has
twice the fundamental magnetic charge~(\ref{Munit}).)
The def\/inition (\ref{taudef2}) allows 
equations~(\ref{Mmn}) and (\ref{WittenEffect}) to be combined into the single complex charge
\begin{gather} \label{QMtau}
Q+iM=g(n_e + n_m \tau)
\end{gather}
and this is a central charge of the supersymmetry algebra.
Semi-classical arguments would then imply modular symmetry (\ref{modulartrans})
on $\tau$ \cite{SW}, but we shall see  
that quantum ef\/fects break full modular symmetry, though 
a subgroup still survives in the full quantum theory.

Classically the Higgs potential is minimised by any constant $\phi$ in the Lie algebra of $SU(2)$
such that $[\phi^\dagger,\phi]=0$, and since 
we can always rotate $\phi$ by a gauge transformation, we can
take $\phi =\frac{1}{2}{\bm a} \sigma_3$ with 
$\sigma_3=
\left(\begin{array}{cc}
1 & 0 \\ 0 & -1 
\end{array}\right)$ the usual Pauli
matrix and ${\bm a}$ a complex constant with dimensions of mass.  The classical vacua are thus
highly degenerate and can be parameterised by ${\bm a}$, a non-zero ${\bm a}$ breaks $SU(2)$ gauge symmetry
down to $U(1)$ (we are still free to rotate around the $\sigma_3$ direction) and $W^\pm$-bosons
acquire a mass proportional to ${\bm a}$, leaving one $U(1)$ gauge boson (the photon) massless.  At the same time
the fermions and the scalar f\/ield also pick up masses proportional to ${\bm a}$, except for the superpartners
of the photon which are protected by supersymmetry and also remain massless.  At the special point ${\bm a}=0$
the gauge symmetry is restored to the full $SU(2)$ symmetry of the original theory.
Supersymmetry protects this vacuum degeneracy so that it is not lifted and is
still there in the full quantum theory.  At low energies, much less than the 
mass ${\bm a}$ for a generic value of ${\bm a}$,
the only relevant
degrees of freedom in the classical theory are the massless $U(1)$ gauge boson and its 
superpartners (quantum ef\/fects modify this for special values of $\tau$).
Dyons with magnetic and electric quantum numbers $(n_m,n_e)$
have a mass ${\cal M}$ that, in ${\cal N}=2$ SUSY, saturates the Bogomol'nyi bound (\ref{BogBound})
which in the conventions used here gives
\begin{gather*}  {\cal M}^2=\frac{2}{g^2}|{\bm a}(Q+iM)|^2\end{gather*}
(some factors of 2 dif\/fer from the original 't Hooft--Polyakov framework).
Motivated by semi-classical modular symmetry
this can be written, using (\ref{QMtau}), as
\begin{gather} \label{classicalmass} {\cal M}^2=2|{\bm a}(n_e + n_m\tau)|^2
=2|n_e {\bm a} +  n_m {\bm a}_D |^2,\end{gather}
where ${\bm a}_D:= \tau {\bm a}$, so the 't Hooft--Polyakov monopole 
has mass ${\cal M}^2=2|{\bm a}_D|^2$.
Under $\Gamma(1)$ $\tau$ transforms as 
\begin{gather*} \gamma(\tau)=\frac{a\tau + b}{c\tau+ d}=
\frac{a\,{\bm a}_D + b\,{\bm a}}{c\,{\bm a}_D+ d\,{\bm a}}
\end{gather*}
so
\begin{gather*} \left(\begin{array}{c} {\bm a}_D \\ {\bm a}  \end{array}  \right)
\quad \rightarrow \quad
\left(\begin{array}{cc} a & b \\ c  & d \end{array}  \right) 
\left(\begin{array}{c} {\bm a}_D \\ {\bm a}  \end{array}  \right)= 
\left(\begin{array}{c} a\, {\bm a}_D+ b\, {\bm a} \\ c\, {\bm a}_D + d\, {\bm a}  \end{array}  
\right).\end{gather*}

The dyon mass ${\cal M}$ is then invariant if 
at the same time $(n_m,n_e)$ is transformed in the opposite way,
\begin{gather*} \left(\begin{array}{c} n_m \\ n_e  \end{array}  \right)
\quad \rightarrow \quad
\left(\begin{array}{cc} d & -c \\ -b  & a \end{array}  \right) 
\left(\begin{array}{c} n_m \\ n_e  \end{array}  \right)= 
\left(\begin{array}{c} d\, n_m- c\, n_e \\ -b\, n_m + a\, n_e \end{array}  
\right).\end{gather*}

In particular, under the analogue of Dirac's electric-magnetic duality:
$\tau\rightarrow \tau_D=-1/\tau$ and $({\bm a}_D,{\bm a})\rightarrow (-{\bm a},{\bm a}_D)$
and $(n_m,n_e)\rightarrow (-n_e,n_m)$. The central charge (\ref{QMtau}) is not invariant under
these transformations
\begin{gather*}
Q+iM\quad\rightarrow\quad \frac{1}{c\tau+d}(Q+iM) \end{gather*}
unless $Q+iM$ is zero or inf\/inite.

When the model is quantised there is a number of very important modif\/ications to the
classical picture:

\begin{itemize}\itemsep=0pt

\item $\tau\rightarrow \tau_D=-1/\tau$ is not a symmetry of the quantum theory.

\item The point ${\bm a}=0$, where full $SU(2)$ symmetry is restored in the classical theory, is
inaccessible in the quantum theory.  The dual strong coupling point ${\bm a}_D=0$ is accessible.
  
\item The ef\/fective low-energy coupling runs as a
function of ${\bm a}$, and since ${\bm a}$ is related to a mass this is
somewhat analogous to the Callan--Symanzik running of the QED coupling.
The ef\/fective gauge coupling only runs for energies greater than $|{\bm a}|$, 
for energies less than this
the running stops and at low energies it becomes frozen at its value $g^2({\bm a})$.
Using standard asymptotic freedom arguments in gauge theory, 
the low energy ef\/fective coupling $g^2({\bm a})$ thus
decreases at large ${\bm a}$ and increases at small ${\bm a}$.  In fact the vacuum angle 
$\theta$ also runs with ${\bm a}$ (as a consequence of instanton ef\/fects \cite{SW,Doreyetal}) 
and this can be incorporated into an ${\bm a}$-dependence of $\tau$, $\tau({\bm a})$.  
At large ${\bm a}$ the imaginary part of $\tau$, $\Im \tau$, becomes large while at
small ${\bm a}$ it is small.  

\item It is no longer true that $\tau={\bm a}_D /{\bm a}$, in the quantum theory
${\bm a}$ and ${\bm a}_D$ are no longer linearly related but instead
\begin{gather*} \tau = \frac{\partial{\bm a}_D}{\partial{\bm a}}.\end{gather*}
The classical mass formula (\ref{classicalmass}) is only true in the quantum theory
in the second form
\begin{gather} \label{quantummass} {\cal M}^2 = \frac{2}{g^2}|n_e{\bm a}+n_m{\bm a}_D|^2.\end{gather}
\end{itemize}

A better, gauge invariant,
parameterisation of the quantum vacua is given by $u={\rm tr}\langle \phi^2\rangle $.  For weak coupling 
(large ${\bm a}$) $u\approx \frac{1}{2}{\bm a}^2$, but 
$\langle \phi^2\rangle \ne\langle \phi\rangle \langle \phi\rangle$ 
for strong coupling (small ${\bm a}$).  By making a few plausible assumptions,
including
\begin{itemize}\itemsep=0pt
\item The low energy ef\/fective action is analytic except for isolated singularities 
(holomorphicity is intimately linked with supersymmetry); 
\item The number of singularities is the minimum possible compatible with stability of the
theory ($\Im \tau >0$),
\end{itemize}
Seiberg and Witten argued \cite{SW} that in
the quantum theory the strong coupling regime
$g^2\approx 0$ is associated not with ${\bm a}=0$ but instead with two points in the complex $u$-plane,
$u=\pm \Lambda^2$ where $\Lambda$ is, by def\/inition, the mass scale at which the gauge coupling
diverges.  Furthermore they found an explicit expression for the 
full low energy ef\/fective action and argued that new massless modes appear at the singular
points $u=\pm \Lambda^2$,
in addition to the photon and its superpartners.  These new massless modes are in fact dyons,
with the magnetic charge coming from topologically non-trivial 
aspects of the classical theory (monopoles).
Since $g\rightarrow \infty$ when $u=\pm \Lambda^2$, $\tau$ is real at these points.
For example the point $\tau=0$ is associated with $u=\Lambda^2$ and the dyons have 
zero electric charge,
they are in fact simple 't Hooft--Polyakov monopoles with $n_m=-1$ 
($M=-\frac{4\pi\hbar}{g}$) and $n_e=0$, also ${\bm a}_D=0$ at this point so ${\cal M}=0$
from (\ref{quantummass}).
The point $\tau=1$ is associated with $u=-\Lambda^2$ and the dyons have $n_e=1$ and $n_m=-1$,
so $Q=0$, and ${\bm a}={\bm a}_D$ so again ${\cal M}=0$.

The full modular group $\Gamma(1)$ is not manifest in the quantum theory, rather
Seiberg and Witten showed that the relevant map is (\ref{modulartrans})
with both $b$ and $c$ constrained to be even.
This is a subgroup of the full modular group, denoted $\Gamma(2)$ in the mathematical literature,
and it is generated by
\begin{gather*} {\cal T}^2: \tau\rightarrow \tau +2 \qquad \hbox{and} \qquad {\cal F}^2: \tau \rightarrow
\frac{\tau}{1-2\tau},\end{gather*}
where ${\cal F}^2={\cal S}^{-1}{\cal T}^2{\cal S}$.\footnote{We def\/ine 
${\cal F}={\cal S}^{-1}{\cal T}{\cal S}:\tau\rightarrow \frac{\tau}{1-\tau}$.}
$u$ itself is invariant under $\Gamma(2)$, but $\tau(u)$ is a multi-branched function of~$u$.

The explicit form of $\tau$ is given in \cite{AlvarezG} as
\begin{gather*} 
\tau =i\frac{K'}{K} + 2n,
\end{gather*}
where $n$ is an integer, $K=K(k)$ and $K'=K(k')$ are standard elliptic integrals with
$k^2=\frac{2}{1+u/\Lambda^2}$ and $k'^2=1-k^2$ (see for example \cite{WW} for properties
of elliptic integrals).  

Seiberg and Witten's $\Gamma(2)$ action commutes with the scaling f\/low
as $u$ is varied.  Taking the logarithmic derivative of $\tau(u)$ with
respect to $u$, and imposing $ad-bc=1$, we see that 
\begin{gather}\label{modularform}
u\frac{d\gamma(\tau)}{du}=\frac{1}{(c\tau +d)^2}u\frac{d\tau}{du}.
\end{gather}
Meromorphic functions $\tau(u)$ satisfying (\ref{modularform}) are well studied in the mathematical literature and are
called modular forms of weight $-2$.

\begin{figure}[t]
\centerline{\includegraphics[width=12cm]{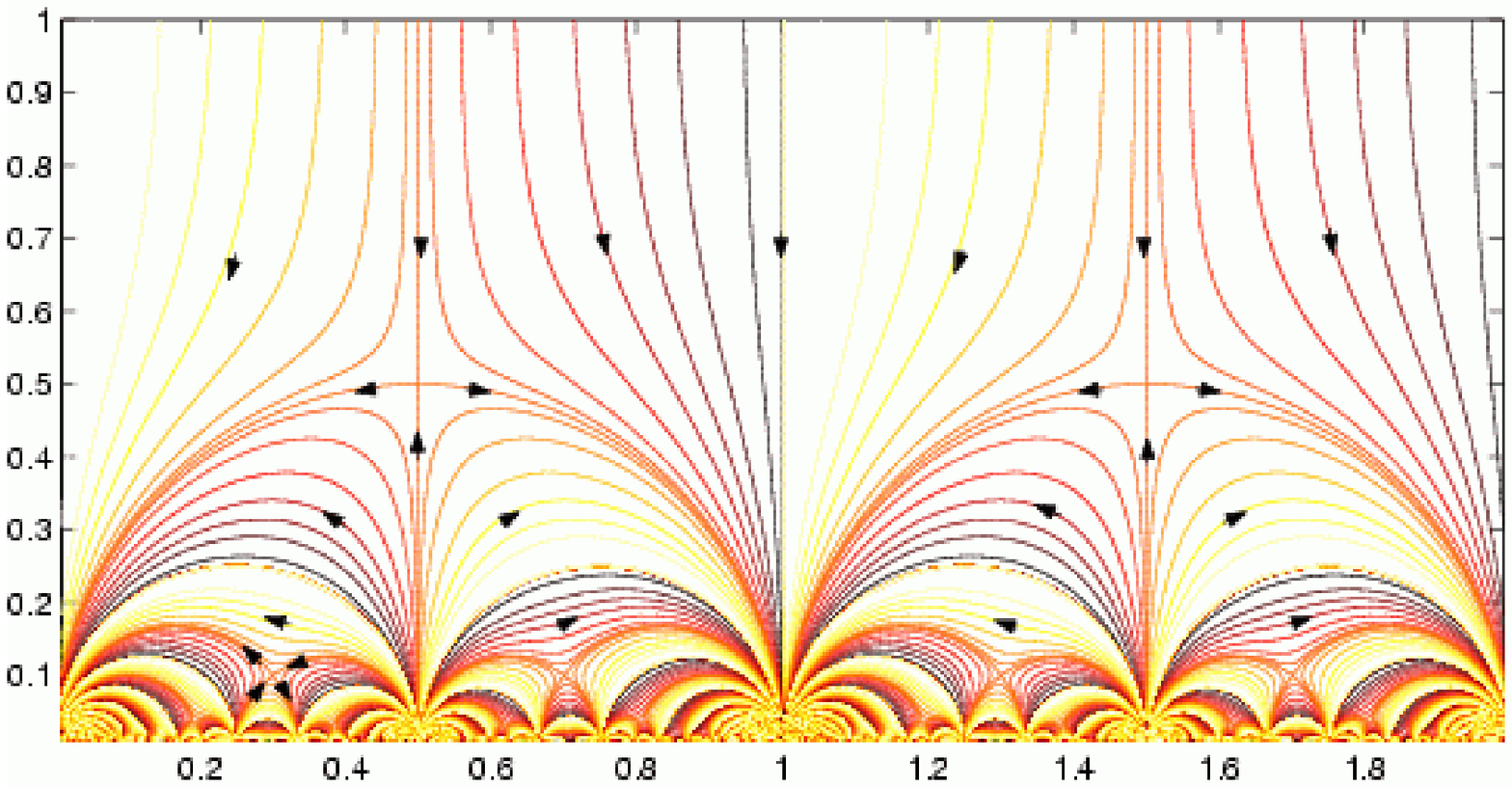}} \caption{\label{fig1}}
\end{figure}

For Seiberg and Witten's expression for $\tau(u)$ 
it was shown in \cite{MinahanNemeschansky,Ritz,Geodesic,Konishi,Konishi+} that
\begin{gather}
-u\frac{d\tau}{du}=
\frac{1}{2\pi i }\left(\frac{1}{\vartheta^4_3(\tau)}+\frac{1}{\vartheta^4_4(\tau)}\right),
\label{betazero}
\end{gather}
where 
\begin{gather*} \vartheta_3(\tau)=\sum_{n=-\infty}^\infty e^{i\pi n^2\tau}\qquad
\hbox{and}\qquad
\vartheta_4(\tau)=\sum_{n=-\infty}^\infty (-1)^ne^{i\pi n^2\tau}\end{gather*}
are Jacobi $\vartheta$-functions\footnote{Def\/initions and
relevant properties of the Jacobi functions can be found in the mathematics literature,
for example \cite{WW}.}.   Scaling functions for ${\cal N}=2$ SUSY Yang--Mills are
also constructed in~\cite{dHoker,dHoker+}.  
  At weak coupling, $g^2\rightarrow 0$, $\tau\rightarrow i\infty$,
$\vartheta_3(\tau)\rightarrow 1$ and $\vartheta_4(\tau)\rightarrow 1$ so 
\begin{gather*} u\frac{d\tau}{du}\approx \frac{a}{2}\frac{d\tau}{da}\rightarrow \frac{i}{ \pi}\end{gather*}
which is the correct behaviour of the ${\cal N}=2$ one-loop $\beta$-function for the
gauge coupling,
\begin{gather}
a\frac{d g}{d a}\approx -\frac{g^3}{ 4\pi^2}. 
\end{gather}
At weak coupling (large ${\bm a}$) ${\bm a}$ is proportional to the gauge boson 
mass and this f\/low can be interpreted as
giving the Callan--Symanzik $\beta$-function 
for the gauge bosons (or equivalently the gauginos because of supersymmetry)
in the asymptotic regime.  
This interpretation is however not valid for f\/inite ${\bm a}$ for two
reasons: f\/irstly the statement that ${\bm a}$ is proportional to the gauge boson mass is only valid
at weak coupling and secondly because (\ref{betazero}) diverges at strong coupling, $g\rightarrow\infty$
where $\tau\rightarrow 0$, \cite{Konishi,Konishi+}.  
The latter dif\/f\/iculty can be remedied by def\/ining a dif\/ferent scaling function
which is still a modular form of weight~$-2$. Seiberg and Witten's expression for~$\tau(u)$
can be inverted to give $u(\tau)$ \cite{Ritz,Geodesic,Konishi,Konishi+}
\begin{gather} \frac{u}{\Lambda^2} =\frac{\vartheta_3^4 + \vartheta_4^4}{\vartheta_3^4 - \vartheta_4^4}\end{gather} 
which is invariant under $\Gamma(2)$ modular transformations \cite{Geodesic,Matone}.
Equation (\ref{betazero}) can therefore be multiplied by any ratio of polynomials in $u$ and we still have
a modular form of weight $-2$.\footnote{This point was stressed in \cite{Ritz}. 
See, for example, theorem 4.3.4 of \cite{Rankin}.}  
 The singularity in $u\frac{d\tau}{du}$ at
$\tau=0$ (where $u=\Lambda^2$) can be removed
by multiplying by $(1-u/\Lambda^2)^k$ with $k$ some positive integer, without introducing any new singularities
or zeros, which is the minimal assumption.  
Indeed the $\beta$-function
at $\tau=0$ can be calculated using one-loop perturbation theory in the dual coupling
$\tau_D=-1/\tau$, \cite{Peskin}, and it is shown in \cite{Nfzero} that one obtains the correct form of 
the Callan--Symanzik $\beta$-function (up to a factor of 2), 
for gauginos at $\tau= i\infty$ and for monopoles at $\tau=0$, using $k=1$,
\begin{gather} -(1-u/\Lambda^2)\Lambda^2\frac{d\tau}{ du}=\frac{1}{\pi i} \frac{1}{\vartheta^4_3(\tau)}.
\end{gather}  
However this is still not correct as there is a second
singularity at $\tau=1$, where $u=-\Lambda^2$.
It is further argued in \cite{Nfzero} that the correct form of 
the Callan--Symanzik $\beta$-function
at $\tau= i\infty$, $\tau=0$ and $\tau=1$, up to a constant factor, is obtained by using
the scaling function
\begin{gather}\label{Nfzerobeta}
-\left(1-\frac{u}{\Lambda^2}\right)\left(1+\frac{u}{\Lambda^2}\right)\frac{\Lambda^4}{u}
\frac{d\tau}{ du} =
\frac{2}{\pi i}\frac{1}{\bigl(\vartheta_3^4(\tau)+\vartheta_4^4(\tau)\bigr)} 
\end{gather}
and this is the minimal choice in the sense that it has the fewest possible singular points.
The f\/low generated by (\ref{Nfzerobeta}) is shown in Fig.~\ref{fig1},
there are f\/ixed points on the real axis at $\tau=q/p$
where the massless dyons have quantum numbers $(n_m,n_e)=(-p,q)$.
Odd $p$
corresponds to attractive f\/ixed points in the IR direction and even $p$ to attractive
f\/ixed points in the UV direction ($p=0$ corresponds to the original weakly coupled theory,
$\tau=i\infty$).  The repulsive singularities at $\tau=\frac{n+i}{2}$ with
$n$ an odd integer occur for $u=0$ and are the quantum vestige of the classical situation were full $SU(2)$
symmetry would be restored. 

One point to note is that, since the scaling function (\ref{Nfzerobeta})
is symmetric under $u\rightarrow -u$, which is equivalent to $\tau\rightarrow\tau+1$, the full
symmetry of the scaling f\/low is slightly larger than $\Gamma(2)$, it is generated by ${\cal F}^2$ and
$\cal T$ and corresponds to matrices $\gamma$ such that $c$ in (\ref{modulartrans}) is even.
This group is often denoted by $\Gamma_0(2)$.

Observe also that there are semi-circular trajectories linking some of the IR attractive f\/ixed points
with odd monopole charge.  
These can all be obtained from the semi-circular 
arch linking $\tau=0$ and $\tau=1$ by the action of some 
$\gamma=\left(
\begin{array}{cc}
a & b \\ c & d 
\end{array}\right)\in\Gamma_0(2)$.  Then 
\begin{gather} \tau_1=q_1/p_1=\gamma(0) = \frac{b}{d} \end{gather} and
\begin{gather} \tau_2=q_2/p_2=\gamma(1)=\frac{a+b}{c+d}\end{gather}
from which $b=\pm q_1$, $d=\pm p_1$ and $q_2=\pm(a+b)$, $p_2=\pm(c+d)$.
Hence
\begin{gather} \label{pqminusqp}
|q_2 p_1 - q_1 p_2| = 1, \end{gather}
since $ad-bc=1$, giving a selection rule for transitions between IR attractive f\/ixed points
as the masses $u$ are varied.  This is clearly related to the Dirac--Schwinger--Zwanziger quantisation
condition (\ref{SZ2}).

When matter in the fundamental representation is included the picture changes in detail, but
is similar in structure \cite{SW2}.  In particular dif\/ferent subgroups of $\Gamma(1)$ appear.
To anticipate the notation let $\Gamma_0(N)\subset \Gamma(1)$ denote the set of matrices
with integral entries and unit determinant
$\gamma=\left(
\begin{array}{cc}
a & b \\ 
c & d 
\end{array}\right)$ 
such that \hbox{$c=0$ mod $N$}
and let $\Gamma^0(N)\subset \Gamma(1)$ denote those with \hbox{$b=0$ mod $N$}.
These are both subgroups of $\Gamma(1)$: 
$\Gamma_0(N)$ being generated by ${\cal T}$ and ${\cal S}^{-1}{\cal T}^N{\cal S}$ while
$\Gamma^0(N)$ is generated by ${\cal T}^N$ and ${\cal S}^{-1}{\cal T}{\cal S}$.

  Now consider ${\cal N}=2$ supersymmetric $SU(2)$ Yang--Mills theory in 4-dimensional Minkowski space
with $N_f$ f\/lavours in the fundamental representation of $SU(2)$.
The low energy ef\/fective action for $0<N_f<4$ was derived in \cite{SW2} ($N_f=4$ is a critical
value for which the quantum theory has $\beta(\tau)=0$ and the theory is conformally invariant).
Because matter f\/ields in the fundamental representation of $SU(2)$
can have half-integral electric charges
it is convenient to re-scale $\tau$ by a factor of two and def\/ine
\begin{gather} \tau'=\frac{\theta}{\pi}+\frac{8\pi i}{g^2}. \end{gather}
Thus 
\begin{gather} \gamma(\tau)=\tau +1 \qquad \Rightarrow \qquad \gamma(\tau')= 2\gamma(\tau)=\tau'+2 \end{gather}
and
\begin{gather} \gamma(\tau) = \frac{\tau}{1-2\tau} \qquad \Rightarrow \qquad 
\gamma(\tau')=2\gamma(\tau)= \frac{\tau'}{1-\tau'}, \end{gather}
so $\Gamma_0(2)$ acting on $\tau$ is equivalent to $\Gamma^0(2)$ acting on $\tau'$.

\begin{figure}[t]
\centerline{\includegraphics[width=11.5cm]{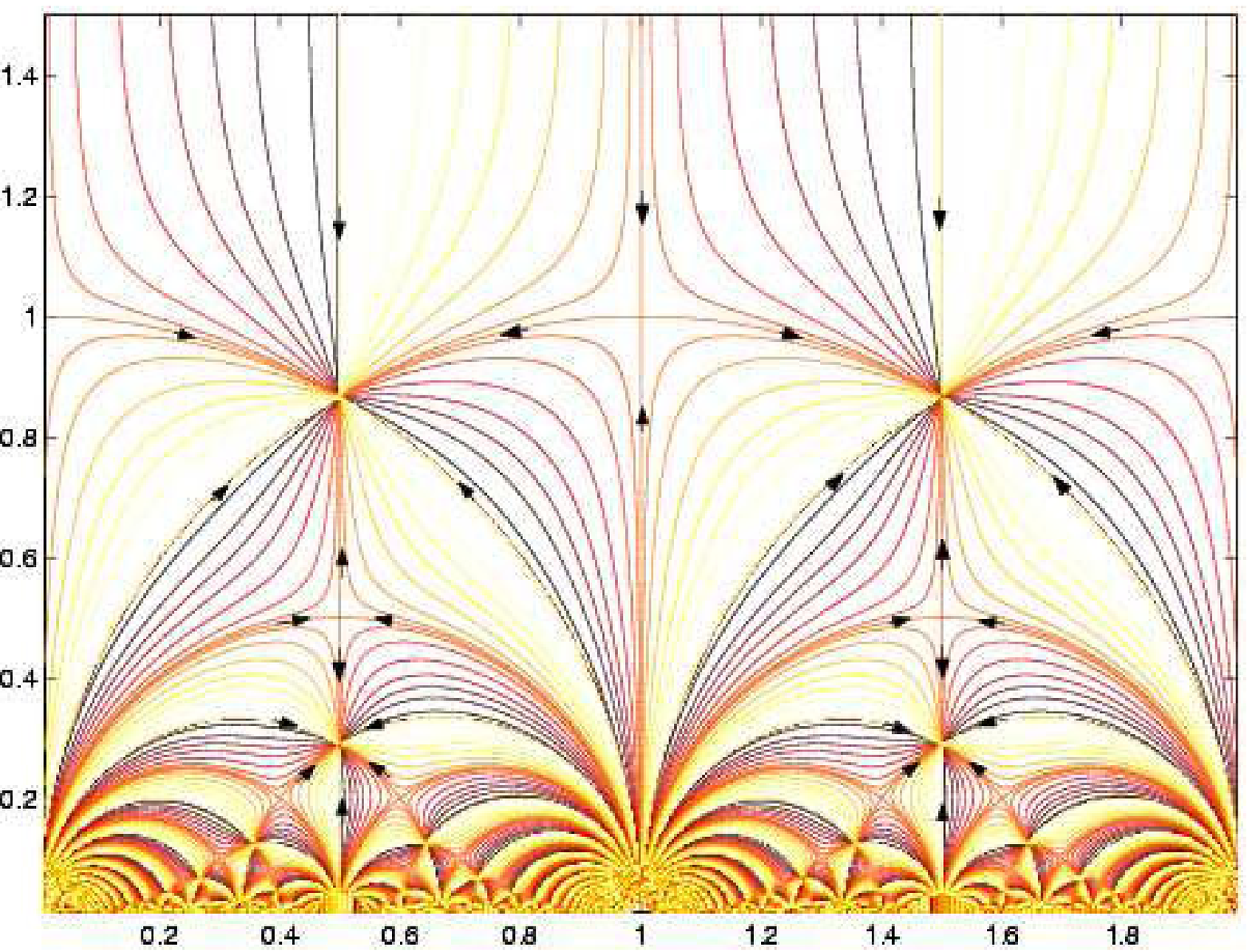}} \caption{\label{fig2}}
\end{figure}

The quantum modular symmetries of the scaling function acting on $\tau'$
are
\begin{gather*}
N_f=0, \qquad \Gamma^0(2), \nonumber \\
N_f=1,  \qquad \Gamma(1), \\
N_f=2,  \qquad \Gamma_0(2), \nonumber \\
N_f=3,  \qquad \Gamma_0(4)\nonumber 
\end{gather*}
and explicit forms of the corresponding modular $\beta$-functions are given in \cite{Matter}.
For $N_f=1$ and $N_f=3$ the group is the same as the symmetry
group acting on the ef\/fective action while for $N_f=0$ and $N_f=2$ it is larger, due to the
${\bf Z}_2$ action on the $u$-plane.  Note the maximal case of $N_f=1$ where the full modular
group is manifest at the quantum level, in this case the duality transformation 
$\tau'\rightarrow -1/\tau'$ is a symmetry and $\tau'=i$ is a f\/ixed point.
The f\/low for $N_f=0$ is shown in Fig.~\ref{fig1} and that for $N_f=1$ is shown in Fig.~\ref{fig2},
the other cases can be found in~\cite{Matter}.

\section[Duality and the quantum Hall effect]{Duality and the quantum Hall ef\/fect}

Modular symmetry manifests itself in the quantum Hall ef\/fect (QHE) in a manner remarkably similar
the that of ${\cal N}=2$ supersymmetric Yang--Mills.  But while ${\cal N}=2$ supersymmetry
is not generally 
believed to have any direct relevance to the spectrum of elementary particles in Nature, the
quantum Hall ef\/fect is rooted in experimental data and its supremely rich structure was not anticipated
by theorists.  The f\/irst suggestion that the
modular group may be related to the QHE was in \cite{WilczekShapere} but, as we shall see below,
the wrong subgroup was identif\/ied in this earliest attempt.

\begin{figure}[t]
\centerline{\includegraphics[width=9cm]{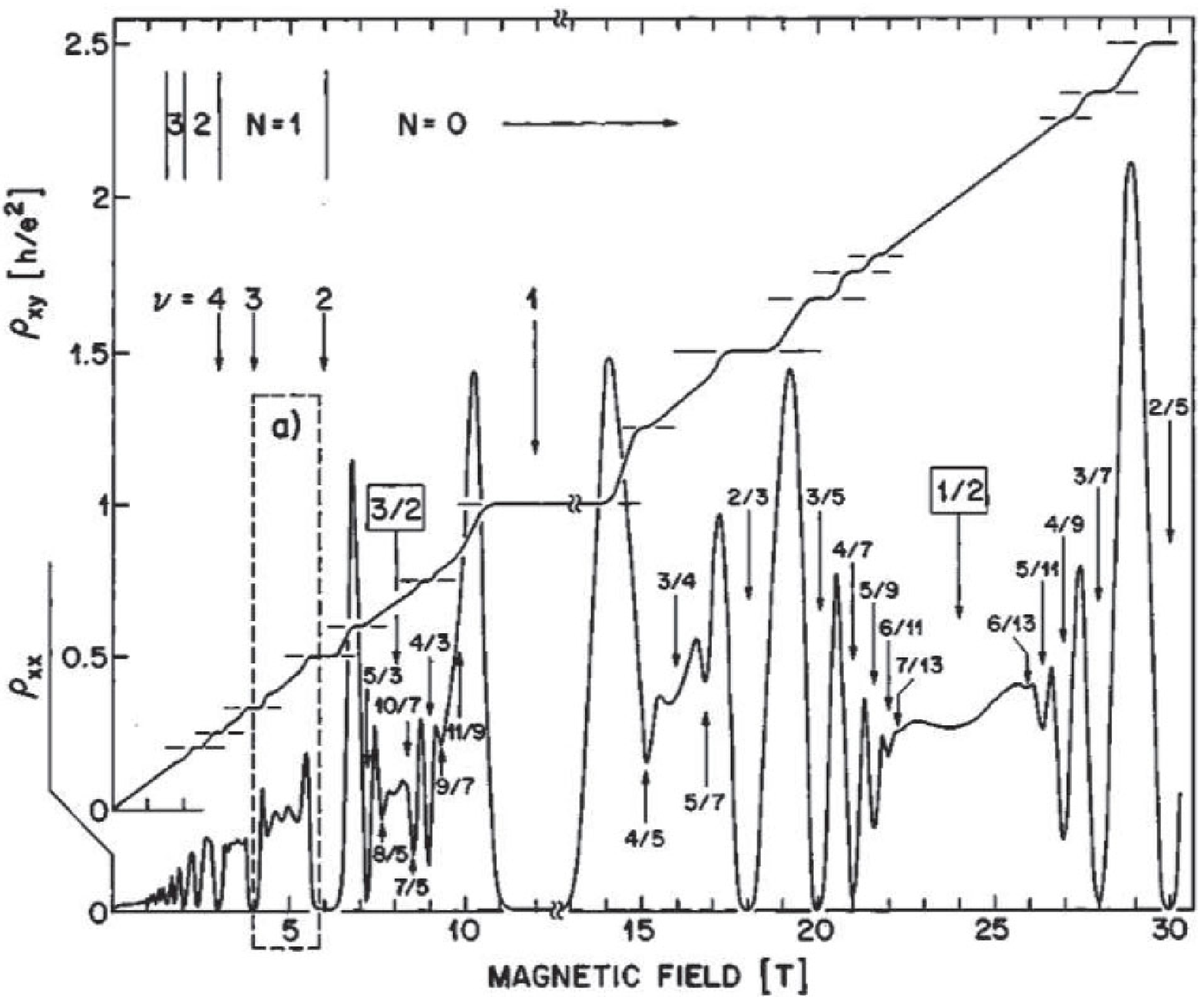}} \caption{\label{fig3}}
\end{figure}

The quantum Hall ef\/fect is a phenomenon associated with 2-dimensional semiconductors
in strong transverse magnetic f\/ields at low temperatures, so that the thermal energy
is much less than the cyclotron energy $\hbar\omega_c$, with $\omega_c$ the cyclotron frequency.  
It requires pure samples
with high charge carrier mobility $\mu$ so that the dimensionless product $B\mu$ is large.
For reviews see e.g.~\cite{Prange,Heinonen,Strong}.  Basically passing a
current $I$ through a rectangular
2-dimensional slice of semi-conducting material requires maintaining a voltage parallel to the
current (the longitudinal voltage $V_L$).  The presence of a magnetic f\/ield $B$ normal to the sample
then generates a transverse voltage (the Hall voltage $V_H$).
Two independent conductivities can therefore be def\/ined: a~longitudinal, or Ohmic, conductivity 
$\sigma_L$ and a transverse, or Hall, conductivity $\sigma_H$, along with the associated resistivities,
$\rho_L$ and $\rho_H$.  
The classical Hall relation is
\begin{gather} \label{CHE} B=e{\tt n}\rho_H\quad\Rightarrow \quad\sigma_HB=-J_e^0 \end{gather}
(with $J_e^0=e{\tt n}$ and  ${\tt n}$ the density of mobile charge carriers) 
and $\sigma_H$ is inversely proportional to $B$
at f\/ixed ${\tt n}$.
In the quantum Hall ef\/fect $\sigma_H$ is quantised as $1/B$ is varied keeping ${\tt n}$ and $T$ f\/ixed
(or varying ${\tt n}$ keeping $B$ and $T$ f\/ixed) and increases in a series
of sharp steps between very f\/lat plateaux.  At the plateaux $\sigma_L$ vanishes and it is non-zero
only for the transition region between two adjacent plateaux.  
In 2-dimensions conductivity has dimensions of $e^2/h$
and, in the f\/irst experiments, \cite{vonKlit} $\sigma_H=n \frac{e^2}{h}$ was 
an integral multiple of $e^2/h$ at the
plateaux (the integer
QHE) though in later experiments \cite{Tsui} it was found that $\sigma_H$ could also be a rational multiple
of $e^2/h$, $\sigma_H=\frac{q}{p}\frac{e^2}{h}$ where $p$ is almost always an odd 
integer (from now in this section on we shall adopt units in which $e^2/h=1$).  
The dif\/ferent quantum Hall plateaux are interpreted as being dif\/ferent phases of a 2-dimensional
electron gas and transitions between the phases can be induced by varying the external magnetic f\/ield,
keeping the charge carrier density constant.  An example of experimental data is shown in Fig.~\ref{fig3},
taken from \cite{Tsui2},
where the Hall resistivity ($\rho_{xy}$) and the Ohmic resistivity ($\rho_{xx}$) for a sample
are plotted as functions of the transverse magnetic f\/ield, in units of $h/e^2$.  The Hall resistivity
is monotonic, showing a series of steps or plateaux, while the Ohmic resistivity shows a series of
oscillations with deep minima, essentially zero, when $\rho_{xy}$ is at a plateaux, and a series
of peaks between the Hall plateaux.

Conductivity is actually a tensor
\begin{gather} J_i=\sigma_{ij}E^j \end{gather} with $\sigma_{xx}$ and $\sigma_{yy}$ the longitudinal conductivities
in the $x$ and $y$ directions and $\sigma_{xy}=-\sigma_{yx}=\sigma_H$ the Hall conductivity associated
with the magnetic f\/ield (for an elementary discussion of the Hall ef\/fect see \cite{LL}).
From now on we shall assume an isotropic medium with $\sigma_{xx}=\sigma_{yy}=\sigma_L$.
Using complex co\"ordinates $z=x+iy$ the conductivity tensor for an isotropic 2-dimensional medium
can be described by a single complex conductivity 
\begin{gather} \sigma:=\sigma_H +i\sigma_L. \end{gather}
Note that Ohmic conductivities must be positive for stability reasons, so $\sigma$ is restricted to
the upper half-complex plane.  The resistivity tensor is the inverse of the conductivity matrix, in
complex notation $\rho_H+i\rho_L=\rho=-1/\sigma$.

In \cite{KLZ} it was argued that the following transformations
\begin{gather} \label{Gamma02sigma} {\cal T}: \sigma\ \rightarrow \ \sigma +1 
\qquad\hbox{and}\qquad
{\cal F}^2: \sigma\ \rightarrow \ \frac{\sigma}{1-2\sigma} \end{gather}
map between dif\/ferent quantum Hall phases of a spin polarised sample.
These transformations 
are known as the Law of Corresponding States for the quantum Hall ef\/fect\footnote{Complex 
conductivities were not used in \cite{KLZ}, but their results are more easily
expressed in the notation used here.}.

The $\cal T$ transformation is interpreted as being due to shifting Landau levels by one, {\it e.g.} 
by varying the magnetic f\/ield keeping ${\tt n}$ f\/ixed.  The philosophy here is that a full Landau level is
essentially inert and does not af\/fect the dynamics of quasi-particles in higher Landau levels.
The number of f\/illed Landau levels is the integer part of the f\/illing factor
$\nu=\frac{{\tt n}h}{eB}$ and
quasi-particles in a partially f\/illed Landau level $n$, with $n$ a non-negative integer,
and f\/illing factor $\nu=n+\delta \nu$, $0\le\delta \nu <1$, will have similar dynamics 
to situations with the same $\delta \nu$ and any other~$n$.  
In particular $n \rightarrow n+1$ should be a symmetry
of the quasi-particle dynamics.  
The situation here is similar to that of the periodic table of the elements were fully f\/illed
electron shells are essentially inert and do not af\/fect the dynamics of electrons in higher shells
-- noble gases would then correspond to exactly f\/illed Landau levels
(the quantum Hall ef\/fect is in better shape than the periodic table of the elements
however because all Landau levels have the same degeneracy whereas
dif\/ferent electron shells have dif\/ferent degeneracies).    
Of course this assumes that there is no inter-level
mixing by perturbations and this can never really be true for all $n$.  To achieve large $n$
for a f\/ixed carrier density ${\tt n}$, for example, requires small $B$ and eventually
$B$ will be so small that that the inter-level gap, which is proportional to $B$, 
will no longer be large compared to thermal energies and/or Coulomb energies.  
 
The $\cal F$ transformation, known as f\/lux-attachment, was used by 
Jain \cite{JainFluxAttachment,JainGoldman}, 
though only for Hall plateaux $\sigma_L=0$, 
as a mapping between ground state wave-functions.
Jain's mapping was associated with modular
symmetry in \cite{SzaboKogan} and with modular transformations of 
partition functions in \cite{Cappelli}.
It is related to the
composite fermion picture of the QHE \cite{Heinonen,JainCompositeFermions,JainCompositeFermions+,Fradkin} where 
the ef\/fective mesoscopic
degrees of freedom are fermionic particles bound to an even number of magnetic f\/lux units, the
operation ${\cal F}^2$ attaches two units of f\/lux to each composite fermion, maintaining their
fermionic nature, in a manner that will be described in more detail below.

The response functions (i.e.\ the conductivities)
in a low temperature 2-dimensional system can be obtained
from a 2+1-dimensional f\/ield theory 
by integrating out all the microscopic physics associated with particles and/or holes 
and incorporating their contribution to the macroscopic physics into ef\/fective coupling constants.  
The classical Hall relation (\ref{CHE})
can be derived from 
\begin{gather} {\cal L}_{\rm eff}[A_0,J_e^0]=\sigma_H A_0 B + A_0 J^0_e, \end{gather}
where $J^0_e$ is the charge density of mobile charges only, not including 
the positive neutralising background of the ions.
The co-variant version of this is 
\begin{gather} \label{SimpleHall} {\cal L}_{\rm eff}[A,J_e]=
\frac{\sigma_H}{2} \epsilon^{\mu\nu\rho} A_\mu \partial_\nu A_\rho  +A_\mu J^\mu_e,\end{gather}
where $\mu,\nu,\rho=0,1,2$ label 2+1 dimensional coordinates.
In linear response theory the Hall conductivity here would be considered to be a response function,
the response of the system to an externally applied electromagnetic f\/ield.
Ohmic conductivity can be included by working in Fourier space $(\omega,{\bf p})$ and 
introducing a frequency dependent electric permittivity.
In a conductor the inf\/inite wavelength electric permittivity has
a pole at zero frequency, in a Laurent expansion
\begin{gather} \label{conductivity}
\epsilon(\omega,0)=-i\frac{\sigma_L}{\omega}+\cdots,\end{gather}
and all of the microscopic physics can be incorporated into response functions that modify
the ef\/fective action for the electromagnetic f\/ield.  In Fourier space, 
in the inf\/inite wavelength, low-frequency limit, the modif\/ication is
\begin{gather}\label{Leff}
 \widetilde{\cal L}_{\rm eff}[A]
\approx
\frac{i\sigma_L}{4\omega}F^2 +
        \frac{\sigma_H}{4}\epsilon^{\mu\nu\rho}A_\mu F_{\nu\rho},\end{gather}
where $F^2=F_{\mu\nu}(-\omega)F^{\mu\nu}(\omega)$ etc.
Of course the full dynamics 
is 3+1 dimensional, unlike the electrons the electric and magnetic f\/ields are not conf\/ined to the
2-dimensional plane of the sample.  What is being written here is the correction to the electromagnetic
action due to the charges in the sample, from which response functions can be read of\/f.
The most relevant correction, in the renormalisation
group sense, is the Chern--Simons term in (\ref{Leff}).
Naively assuming that there are no large anomalous dimensions arising from
integrating out the microscopic degrees of freedom, the next most relevant term would be the
usual Maxwell term.  It is an assumption of the analysis in \cite{KLZ} that other terms,
higher order in $F$ and its derivatives, are not relevant. 

Note that the right hand side of (\ref{Leff}) is not real, 
an indication of the dissipative nature of Ohmic resistance,
and non-local in time, again a feature of a conducting medium.  
Also we have used a relativistic notation and $F^2$ should really be split into
${\bf E}\cdot {\bf E}$ and ${\bf B}\cdot {\bf B}$ with dif\/ferent coef\/f\/icients (response functions).  
In the ${\bf p}=0$,
low-frequency limit of a conductor however, both response functions behave as $1/\omega$, the ratio
is a constant and a relativistic notation can be 
used\footnote{For f\/inite $\omega$ and/or non-zero ${\bf p}$ this would not be the case as the
response functions for ${\bf E}$ and ${\bf B}$ would be dif\/ferent functions in general.}. 
Chern--Simons theories of the QHE have been considered by a number of 
authors \cite{KLZ,Girvin,Anyons2,ZHK,Frohlich,Bal,FradkinCS1,FradkinCS2}.
The presence of the $F^2$ term has been analysed from the general point of view of 3-dimensional
conformal f\/ield theory in~\cite{Witten,Leigh,Zucchini,Lee}.

\begin{figure}[t]
\centerline{\includegraphics[width=6cm]{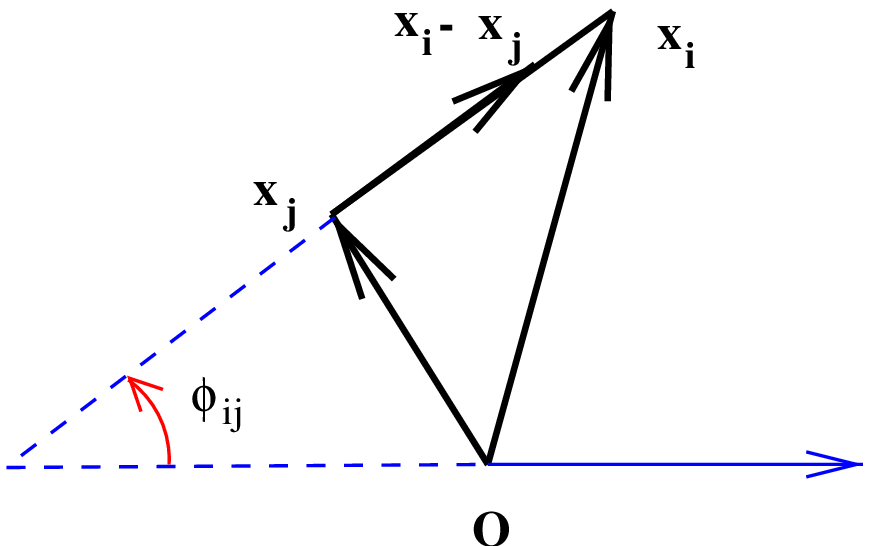}} \caption{\label{fig4}}
\end{figure}

For strong magnetic f\/ields however (\ref{Leff}) is not small and linear response theory
cannot be trusted.  This problem can be evaded by introducing what is called the 
``statistical gauge f\/ield'',~$a_\mu$, \cite{Fradkin,Anyons2,Anyons1,GMcD}, and then following the
analysis of \cite{KLZ}. 
Consider a sample of material with $N$ charge carriers with wave-function $\Psi({\bf x}_1,\ldots,{\bf x}_N)$.
Perform a gauge transformation
\begin{gather} \label{statphase}\Psi({\bf x}_1,\ldots,{\bf x}_N) \ \rightarrow\ 
\Psi'({\bf x}_1,\ldots,{\bf x}_N) = e^{i\vartheta
(\sum\limits_{i <  j}
\phi_{ij})}\Psi({\bf x}_1,\ldots,{\bf x}_N),\end{gather}
where $\phi_{ij}$ is the angle related to the positions of particle $i$ and particle $j$
as shown in Fig.~\ref{fig4},
in terms of complex coordinates, ${\bf z}=x+iy$, 
$e^{i\phi_{ij}}=\frac{{\bf z}_i - {\bf z}_j}{|{\bf z}_i - {\bf z}_j|}$.
$\vartheta$ is a constant parameter and can
change the statistics of the particles: for example if the particles are fermions, so that
$\Psi$ is anti-symmetric under interchange of any two particles $i$ and $j$, i.e.\ when
$\phi_{ij} \rightarrow \phi_{ij}+\pi$, then $\Psi'$ is
again anti-symmetric if $\vartheta=2k$ is an even integer, $\Psi'$  becomes symmetric if
$\vartheta=2k+1$ is an odd integer.

For co-variant derivatives, $({-i\hbar\nabla-eA})_\mu$ acting on $\Psi$ becomes   
$({-i\hbar\nabla-eA-a})_\mu$ acting on~$\Psi'$ under (\ref{statphase}) where 
\begin{gather} a_\alpha({\bf x}_i)=\hbar\vartheta\sum_{j\ne i}\nabla^{(i)}_\alpha\phi_{ij}
=-\hbar\vartheta\sum_{j\ne i}\epsilon_{\alpha\beta}
\frac{({\bf x}_i-{\bf x}_j)^\beta}{|{\bf x}_i-{\bf x}_j|^2}, \qquad (\alpha,\beta=1,2).
\end{gather} 
As long as the particles have a repulsive core, so ${\bf x}_i\ne{\bf x}_j$
for $i\ne j$, the potential $a$ is a pure gauge, but if the particles can coincide
there is a vortex singularity in $a$ and
\begin{gather} \label{statgauge}
\epsilon^{\beta\alpha}\nabla^{(i)}_\beta a_\alpha({\bf x}_i)=
h\vartheta \sum_{j}\delta^{(2)}({\bf x}_i-{\bf x}_j)\rightarrow
h\vartheta{\tt n}({\bf x}_i), \end{gather}
the last expression being the continuum limit of the discrete particle distribution.
The statistical gauge f\/ield $a_\mu$
was introduced in the context of
``anyons'' in \cite{Anyons2,Anyons1}.  Its relation with the charge carrier 
density (\ref{statgauge}) was pointed out in \cite{GMcD}, where similarities between the
statistics parameter~$\vartheta£$ and the vacuum angle QCD were noted and
it was observed the binding of vortices to particles
is a $2+1$-dimensional version of 't Hooft's notion of oblique conf\/inement in QCD~\cite{tHooft},
in which a condensate of composite objects occurs.

The constraint (\ref{statgauge}) can be encoded
into the dynamics, by introducing a Lagrange multip\-lier~$a_0$, and then made co-variant by 
modifying (\ref{SimpleHall}) to read
\begin{gather} \label{LeffAaJ}{\cal L}_{eff}[A,a,J_e]=
\frac{s}{2} \epsilon^{\mu\nu\rho} A_\mu \partial_\nu A_\rho  +\left(A+ \frac{a}{e}\right)_\mu J^\mu_e
-\frac{1}{2\vartheta e^2} \epsilon^{\mu\nu\rho}a_\mu\partial_\nu a_\rho,
\end{gather}
(using units with $\frac{e^2}{h}=1$) where $J^\mu_e$ is the current generated by the matter f\/ields $\Psi'$.
Note that the coef\/f\/icient of the $A_\mu$ Chern--Simons term has been changed from $\sigma_H$ to a new
parameter~$s$, this is because quantum ef\/fects modify the Hall coef\/f\/icient and identifying $\sigma_H$ with
$s$ is premature at this stage.  
Following \cite{KLZ} it is now argued that integrating out $\Psi'$ in (\ref{LeffAaJ}) 
to get an ef\/fective action for $A$ and $a$ can only produce terms that depend on the 
combination $A':=A+a/e$ and gauge invariance restricts the allowed terms\footnote{Note that Euclidean
signature was used in \cite{KLZ} while we use Minkowski signature here.}.  Def\/ine the
f\/ield strength for $A'$ as usual
\begin{gather} F'_{\mu\nu}=\partial_\mu A'_\nu - \partial_\nu A'_\mu
\end{gather}
and that of $a$ as
\begin{gather} f_{\mu\nu}=\partial_\mu a_\nu - \partial_\nu a_\mu.\end{gather}
Then the ef\/fective action, in Fourier space in the long-wavelength ${\bf p}=0$ limit, will be of the form
\begin{gather} \label{LeffAa}
\widetilde{\cal L}_{eff}[A,a]=\frac{s}{4}\epsilon^{\mu\nu\rho}A_\mu F_{\mu\nu}-\frac{\Pi_L(\omega)}{4}F^{'2}+
        \frac{\Pi_H(\omega)}{4}\epsilon^{\mu\nu\rho}A'_\mu F'_{\nu\rho}
-\frac{1}{4\vartheta e^2} \epsilon^{\mu\nu\rho}a_\mu f_{\nu\rho},\end{gather}
where the response functions $\Pi_L$ and $\Pi_H$ cannot be calculated but are {\it assumed} to
give conductivities in the $\omega\rightarrow 0$ limit, i.e.\ it is assumed that $\Pi_L$ has a pole 
while $\Pi_H$ is f\/inite at $\omega=0$,
\begin{gather} i\omega\Pi_L(\omega)\ \mathop{\longrightarrow}_{\omega\rightarrow 0}\ \sigma_L,\qquad 
\Pi_H(\omega)\ \mathop{\longrightarrow}_{\omega\rightarrow 0}\ \sigma_H.\end{gather}
Integrating out $\Psi'$ will in general produce many more terms in the ef\/fective action than
shown in (\ref{LeffAa}), higher order
powers in $F'$ and its derivatives, with unknown coef\/f\/icients, but it is an 
assumption of the analysis that these are less relevant, in the
renormalisation group sense, than the $\Pi_L$ and $\Pi_H$ terms exhibited here.
This does require a leap of faith however -- while it is certain that there are no anomalous dimensions
associated with the Chern--Simons term, $\epsilon A' F'$ since it is topological, it is by no
means obvious that integration of $\Psi'$ will not produce
large anomalous dimensions associated with operators like $(F'^{2})^2$,
for example, that might render them more relevant that $F'^{2}$.  It is an assumption in \cite{KLZ} that this
does not happen.  Note however that it {\it has} been assumed that there is a large anomalous dimension
associate with $F'^{2}$ -- this operator is not naively marginal in 2+1-dimensions, but the assumption
of a pole in $\Pi_L$ renders it marginal.  The ef\/fective action (\ref{LeffAa}) is actually conformal 
\cite{Witten}.  One further modif\/ication of the ef\/fective action used in \cite{KLZ} is
to allow for a change in the ef\/fective charge of the matter f\/ields $\Psi'$, $e^*=\eta e$.  
This is easily incorporated into (\ref{LeffAa}) by redef\/ining $A'$ to be $A'=\eta A + a/e$ 
everywhere (the units are not af\/fected, we still have $h=e^2$). 

The ${\cal F}^2$ and ${\cal T}$ transformations 
in (\ref{Gamma02sigma}) can now be derived by integrating $a$ out of the ef\/fective action (\ref{LeffAa}),
which is quadratic in $a$ and so the integration is Gaussian.  This leads to a new ef\/fective action
for $A$ alone, with dif\/ferent conductivities, 
\begin{gather} {\widetilde{\cal L}_{\rm eff}}^\prime [A]=\frac{i\sigma_L'}{4\omega}F^2+
        \frac{\sigma_H'}{4}\epsilon^{\mu\nu\rho}A_\mu F_{\nu\rho}\end{gather}
where $\sigma'_L$ and $\sigma_H'$ are best
expressed in terms of the complex conductivity  
$\sigma'=\sigma_H' + i\sigma_L'$
as a~fractional linear transformation
\begin{gather} \sigma'=\frac{a\sigma + b}{c\sigma+d}\end{gather}
and one f\/inds  \cite{KLZ}
\begin{gather} \label{KLZgamma}
\left(\begin{array}{cc} 
a & b \\ 
c & d 
\end{array}\right)=
\frac{1}{\eta} \left(\begin{array}{cc} 
\eta^2 + s\vartheta & s \\ 
\vartheta & 1 
\end{array}\right)
:=\gamma(\eta,s,\vartheta).\end{gather}
The analysis of was taken further in 
\cite{Modular1} and extended beyond the regime of linear response in~\cite{Modular2}.

Note that $\det\gamma=1$ so the most general transformation, for arbitrary $s$, $\vartheta$ and $\eta$,
is an element of $Sl(2;{\bf R})$.  However a general element is not a symmetry, for example if $\vartheta$ 
is not an integer then the two conductivities describe charge carriers which are anyons with
dif\/ferent statistics.  Also if $\vartheta$ is an odd integer we have mapped between bosonic and
fermionic charge carriers and this is not a symmetry, though this was the map described
in the analysis in
\cite{KLZ} where $\sigma$ was associated with a bosonic system, which could form
a Bose condensation, and $\sigma'$ was associated with a quantum Hall system
with fermionic charge carriers (there is no suggestion here that a~supersymmetric theory
necessarily underlies the quantum Hall ef\/fect).  

Quantum mechanics requires that the parameters
$\eta$, $s$ and $\vartheta$ in (\ref{KLZgamma}) must necessarily be quantised for the map to be a symmetry.
For example $\eta=1$, when the charge carriers are pseudo-particles with $e^*=e$, 
with $\vartheta=0$
gives 
\begin{gather} \gamma(1,s,0)=\left(\begin{array}{cc} 
1 & s \\ 
0 & 1 
\end{array}\right)\end{gather}
and we only expect this to be a symmetry when $s$ is an integer (Landau level addition).  
In particular $\gamma(1,1,0)$ gives the ${\cal T}$ transformation of (\ref{Gamma02sigma}).
The f\/lux attachment transformation, ${\cal F}^2$ in (\ref{Gamma02sigma}), follows from the fact that
$\vartheta=2k$ should be a symmetry for any integral $k$, with $\eta=1$ and $s=0$.  In particular
\begin{gather} \gamma(1,0,-2)=\left(\begin{array}{cc} 
1 & 0 \\ 
-2 & 1 
\end{array}\right)\end{gather}
gives ${\cal F}^2$.

We can now map between dif\/ferent Hall plateaux by combining the operations of adding statistical f\/lux
and Landau level addition.
For example if we start from a plateau with $\sigma=1$  
and perform a (singular) gauge transformation with $\vartheta = 2$, which is ${\cal F}^{-2}$, then we transform to 
a new phase with $\sigma=1/3$.

Note that a general transformation requires $\eta\ne 1$, although it is always rational.
For example the Jain series of fractional quantum Hall 
states\footnote{See for example the article by Jain and Kamilla in \cite{Heinonen}.} 
can be generated by
starting from $\sigma=0$
and mapping to Landau level $q$ using ${\cal T}^q$, where $q$ is a positive integer. 
Now add $2k$ units of statistical f\/lux using ${\cal F}^{-2k}$, the resulting
transformation is
\begin{gather*} {\cal F}^{-2k}{\cal T}^q=\left(\begin{array}{cc} 
1 & 0 \\ 
2k & 1 
\end{array}\right)
\left(\begin{array}{cc} 
1 & q \\ 
0 & 1 
\end{array}\right)=
\left(\begin{array}{cc} 
1 & q \\ 
2k  & 2 k q+1 
\end{array}\right)=
\gamma\left(\frac{1}{2kq+1},\frac{q}{2kq+1},\frac{2k}{2kq+1}\right).
\end{gather*}
The new conductivity has 
\begin{gather*} \sigma'=\frac{q}{p} \qquad \hbox{with}\qquad p=2kq+1 
\qquad \hbox{and}\qquad \eta=\frac{1}{p},
\end{gather*}
so the pseudo-particles carry not only $2k$ vortices of statistical f\/lux but also electric charge
\begin{gather*} e^*=\frac{e}{p}
\end{gather*}
with $p$ odd,  i.e.\ {\it fractional charge}.
This is precisely analogous to the Witten ef\/fect described in Section~2.
Fractional charges of $e/3$ for $q=k=1$ were predicted in \cite{LaughlinFrac} and
have been conf\/irmed by experiment \cite{Picciotto,Saminadayar}.

The model of the Hall plateaux espoused in \cite{KLZ} is that one has charge carriers which
are fermions (electrons or holes) interacting strongly with the external f\/ield.  By attaching
an odd number $2k+1$ of statistical f\/lux units to each fermion the resulting composite particles
are bosons.  By choosing $k$ appropriately it can be arranged that the 
ef\/fect of the external magnetic f\/ield is almost cancelled by the statistical gauge f\/ield and the composite
bosons behave almost as free particles.  Being bosons they can condense to form a superconducting phase,
with a mass gap, and this explains the stability of the quantum Hall plateaux for the original fermions.
This map between bosons and fermions is of course not a symmetry, nevertheless it is 
a useful way of looking at the physics.
An alternative view is Jain's composite fermion picture \cite{Heinonen,JainCompositeFermions,JainCompositeFermions+,Fradkin}, 
which is a symmetry.  
Both the composite fermion and the composite boson picture are useful, but in either case 
fermions lie at the heart of the physics, as indicated by the antisymmetry of the Laughlin ground state
wave-functions.

The transformations (\ref{Gamma02sigma}) 
acting on the complex conductivity map between dif\/ferent quantum Hall phases, 
clearly they generate the group $\Gamma_0(2)$.
As mentioned above the relevance of the modular group to the QHE was anticipated by Wilczek and 
Shapere \cite{WilczekShapere}, though these authors focused on a dif\/ferent subgroup of $\Gamma(1)$, one
generated by 
\begin{gather} \label{GammaTheta} {\cal S}:\sigma\rightarrow-1/\sigma 
\qquad \hbox{and} \qquad{\cal T}^2:\sigma\rightarrow\sigma+2
\end{gather}
(denoted by $\Gamma_\theta$ here and by $\Gamma_{1,2}$ in \cite{WilczekShapere}) and 
the experimental data on the QHE do not bear this
out.  For example $\Gamma_\theta$ has a f\/ixed point at $\sigma=i$ there is no such f\/ixed point
in the experimental data for the
quantum Hall ef\/fect.  We shall return to $\Gamma_\theta$ below in the context of 2-dimensional 
superconductors.  
L\"utken and Ross \cite{LutkenRoss1} observed,
even before \cite{KLZ}, that the quantum Hall phase diagram in the complex 
conductivity plane bore a striking resemblance
to the structure of moduli space in string theory, at least for toroidal geometry, 
and postulated that $\Gamma(1)$ was relevant to the
quantum Hall ef\/fect.  In \cite{LutkenRoss2, Lutken,Lutken+} the subgroup $\Gamma_0(2)$ was identif\/ied
as being one that preserves the parity of the denominator and therefore likely to be associated
with the robustness of odd denominators in the experimental data.  

Among the assumptions that go into the derivation of (\ref{Gamma02sigma}) from (\ref{Leff}) 
are that the temperature is suf\/f\/iciently low and that the sample is suf\/f\/iciently pure, 
but unfortunately the
analyses in \cite{LutkenRoss1,LutkenRoss2,KLZ,Modular2} are
unable to quantify exactly what ``suf\/f\/iciently'' means.
Basically one must simply assume that (\ref{Leff}) contains the most relevant terms for obtaining
the long wavelength, low frequency response functions,
but obviously this is not always true even at
the lowest temperatures, for example it is believed that, at zero temperature,
a Wigner crystal will form for f\/illing fractions
below about 1/7 and (\ref{Leff}), being rotationally invariant, 
cannot allow for this\footnote{Fractions as low as 1/9 have been reported at f\/inite temperature \cite{Oneover9}.}.

Of course $\Gamma_0(2)$ is not a symmetry of all of the physics, after all the conductivities dif\/fer
on dif\/ferent plateaux, nevertheless it is a symmetry of some physical properties.
The derivation of the $\Gamma_0(2)$ action in \cite{KLZ} required performing a Gaussian
integral about a f\/ixed background, dif\/ferent backgrounds give dif\/ferent initial conductivities,
but in each case the dynamics contributing to the f\/luctuations are the same and the
dynamics of the f\/inal system have the same form as that of the initial one.
This motivates the suggestion \cite{Modular0,Taniguchi,BurgLut,BurgLut+} that the scaling f\/low, which
is governed by the f\/luctuations, should commute with $\Gamma_0(2)$, i.e.\ although
$\Gamma_0(2)$  is not a~symmetry of all the physics it is a symmetry of the scaling f\/low.
Physically the scaling f\/low of the QHE can be viewed as arising from changing
the electron coherence length $l$,  e.g.\ by varying the temperature $T$ with $l(T)$
a monotonic function of $T$ \cite{Khmelnitskii,PruiskenScaling}. Def\/ine a scaling function by
\begin{gather*}
\Sigma(\sigma,\bar\sigma):=l\frac{d\sigma}{dl}.
\end{gather*}

In general one expects $\sigma$
to depend on various parameters, such as the temperature $T$, 
the external f\/ield $B$, the charge carrier density $\tt n$ and the impurity density ${\tt n}_I$.
If ${\tt n}$ and ${\tt n}_I$ are f\/ixed then $\sigma(B,T)$ becomes a function of $B$ and $T$ only.
The scaling hypothesis of \cite{PruiskenScaling} suggests that, at low temperatures, $\sigma$
becomes a function of a single scaling variable.

Now for any 
$\gamma\in\Gamma(1)$,  $\gamma(\sigma)=\frac{a\sigma +b}{c\sigma +d}$
with $ad-bc=1$, so
\begin{gather*} \Sigma\bigl(\gamma(\sigma),\gamma(\bar\sigma)\bigr)=
\frac{1}{(c\sigma +d)^2}\Sigma(\sigma,\bar\sigma), 
 \end{gather*}
this is always true, by construction, and does not represent a symmetry.
 
Demanding that $\Gamma_0(2)$ is a symmetry of the QHE f\/low we immediately get very strong predictions
concerning quantum Hall transitions.
Firstly one can show that any f\/ixed point of~$\Gamma_0(2)$, with $\sigma_{xx}>0$, must be
f\/ixed point of the scaling f\/low (though not {\it vice versa}), \cite{Modular0}.
To see this assume that the scaling f\/low commutes with the action of $\Gamma_0(2)$ and
let $\sigma_*$ be a~f\/ixed point of $\Gamma_0(2)$, i.e.\ there exists a $\gamma\in\Gamma_0(2)$
such that $\gamma(\sigma_*)=\sigma_*$.
If $\sigma_*$ were not a scaling f\/ixed point, we could move to an
inf\/initesimally close point 
$\delta f(\sigma_*) \neq \sigma_*$
with an inf\/initesimal scaling transformation, $\delta f$.  Assuming $\gamma\bigl(\delta f(\sigma_*)\bigr)=
\delta f\bigl(\gamma(\sigma_*))=\delta f(\sigma_*)$
then implies that
$\delta f(\sigma_*)$ is also left invariant by $\gamma$. But, for $\Im\sigma>0$
and f\/inite, the 
f\/ixed points of $\Gamma_0(2)$
are isolated and there is no other f\/ixed point of $\Gamma_0(2)$
inf\/initesimally close to $\sigma_*$.  Hence
$\delta f(\sigma_*)=\sigma_*$ and~$\sigma_*$ must be a scaling f\/ixed point.
Fixed points $\sigma_*$ of $\Gamma_0(2)$
with $\Im \sigma>0$ are easily found by solving the equation $\gamma(\sigma_*)=\sigma_*$: there is
none with $\Im \sigma>1/2$: there is one at $\sigma_*=\frac{1+i}{2}$, 
indeed a~series with $\sigma_*=\frac{n+i}{2}$ for all integral $n$,
and of course the images of these under $\Gamma_0(2)$, leading to a fractal structure as one approaches
$\Im\sigma=0$.  Many of these f\/ixed points have been seen in experiments~\cite{Modular0}.

Experiments show that f\/ixed points between plateaux correspond to second order phase transitions between
two quantum Hall phases and modular symmetry implies that scaling exponents
should be the same at each f\/ixed point, 
a phenomenon known as {\it super-universality} \cite{SUnivWTPP,SUnivEngel}.  
The idea here is that the transition between two quantum Hall plateaux
is a quantum phase transition \cite{QPT} and, at low temperatures, $\sigma$
becomes a function of the single scaling variable $\frac{\Delta B}{T^\kappa}$ with a~scaling exponent $\kappa$,
where $\Delta B=B-B_c$ with $B_c$ the magnetic f\/ield at the critical point. 
While there is evidence for super-universality 
with $\kappa\approx 0.44 $ \cite{SUnivWTPP, Shaharetal}, 
there are some experiments which seem to violate it~\cite{HilkeViolation}.  This may be
due to the suggestion that there exists a marginal operator at the critical point~\cite{Zirnbauer},
but an alternative explanation, that the scaling function might be
a~modular form, is proposed in~\cite{Analytic}.

A second striking consequence of $\Gamma_0(2)$ symmetry of the scaling f\/low is 
a selection rule for quantum Hall transitions which can be derived in exactly the same way as
in the discussion around (\ref{pqminusqp}) in Section~3.  Given that the integer transition
$\sigma:2\rightarrow 1$ is observed we conclude that $\gamma(2)\rightarrow \gamma(1)$ should
also be possible.  Let $\gamma(1)=\frac{q_1}{p_1}$ and $\gamma(2)=\frac{q_2}{p_2}$  
where $q_1$ and $p_1$ are mutually prime, as are $q_2$ and $p_2$.  
Then, with $\gamma=\left(\begin{array}{cc} a & b \\ c & d  \end{array} \right)$, 
\begin{gather} \frac{q_1}{p_1}=\frac{a+b}{c+d}\qquad \hbox{and} \qquad \frac{q_2}{p_2}=\frac{2a+b}{2c+d}\end{gather} 
so,
\[
q_1=\pm(a+b),\qquad p_1=\pm(c+d), \qquad q_2=\pm(2a+b) \qquad \hbox{and} \qquad p_2=\pm(2c+d).
\]
Since $ad-bc=1$ we obtain the selection rule \cite{Modular0}
\begin{gather}
|q_2 p_1 - q_1 p_2| =1.
\end{gather}
This selection rule is well borne out by the experimental data in Fig.~\ref{fig3}.  
In spin-split samples any 
two adjacent well-formed plateaux, with no unresolved sub-structure between them, obey this 
rule\footnote{Note the point labelled $11/9$ in Fig.~\ref{fig3}
is not compatible with the adjacent plateau at $9/7$ according to this selection rule. 
However $11/9$  is merely a kink in the f\/low which has
probably been identif\/ied wrongly.  The selection rule implies that this kink is most likely be
the precursor of the $14/11$ plateau and that there is unresolved structure between $14/11$ and the
integer plateau at $1$.}. 

\begin{figure}[t]
\centerline{\includegraphics[width=7cm]{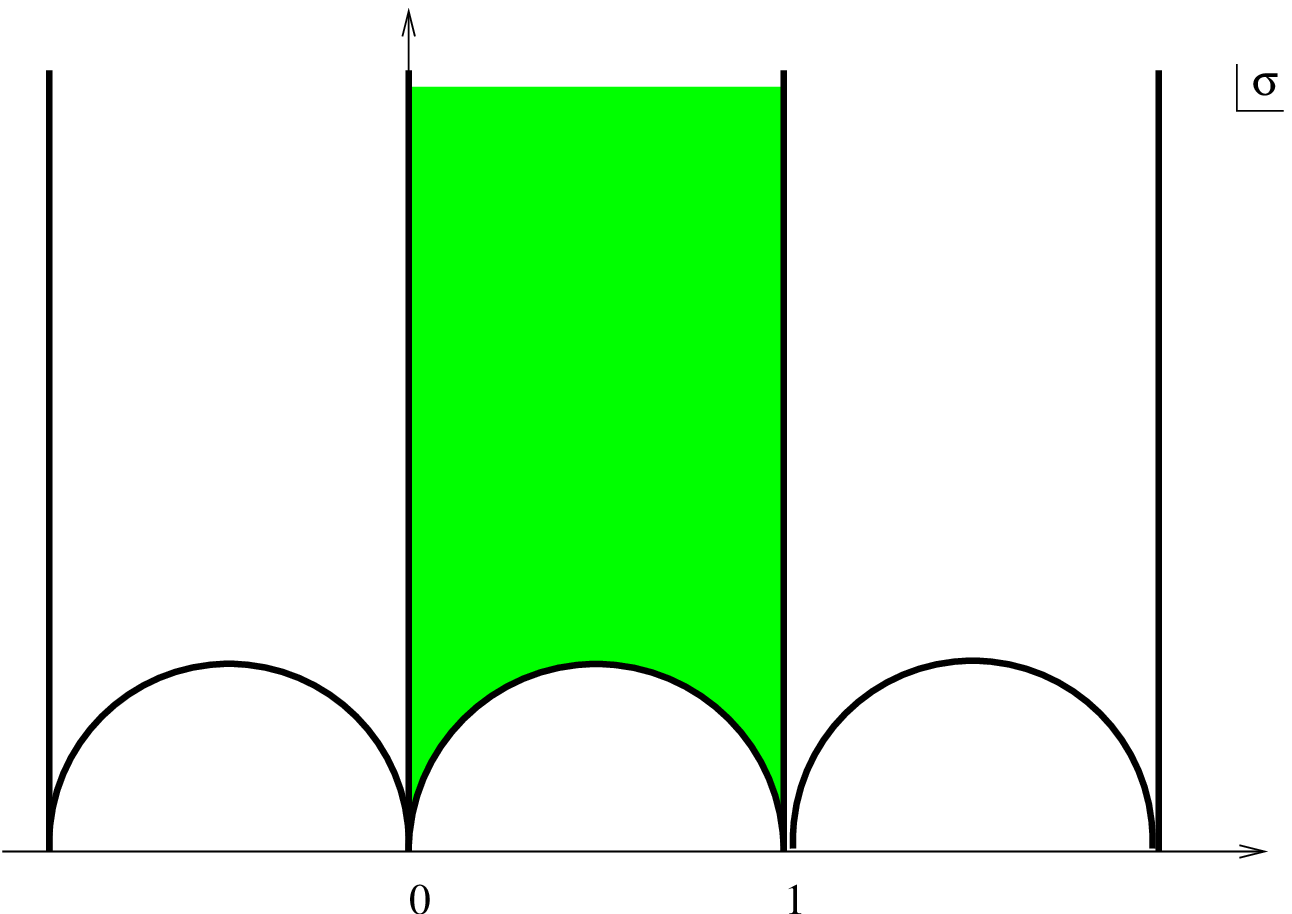}} \caption{\label{fig5}}
\end{figure}

If we further assume that the {\it only} f\/ixed points of the scaling f\/low are the
f\/ixed points of $\Gamma_0(2)$ then, with a few extra reasonable assumptions, 
the topology of the f\/low diagram is completely determined and it is exactly the same as
in Fig.~\ref{fig1} for ${\cal N}=2$ SUSY. 
The whole upper-half plane can be covered by starting with the inf\/inite strip
of width 1 above the semi-circular arch spanning $0$ and $1$ in Fig.~\ref{fig1} (called the
fundamental domain of $\Gamma_0(2)$ and shown in Fig.~\ref{fig5},~\cite{Rankin}) 
and acting on it with elements of 
$\Gamma_0(2)$.  IR f\/ixed points have $\tau=q/p$ corresponding to fermionic charge carriers 
which can be viewed, by turning \cite{KLZ} around, as 
bosons with $p$ units of statistical f\/lux attached, where $p$ is odd.

Guided by experiment, the reasonable assumptions are \cite{Analytic}: 
\begin{itemize}\itemsep=0pt
\item Rational
numbers $q/p$ with odd $p$ are attractive, since they are experimentally so stable;
\item  The
f\/low comes down vertically from the points at $\sigma=\sigma_H+i\infty$ i.e.\
that $\sigma_H$ does not f\/low at high temperatures (weak coupling \cite{Analytic,PruiskenSigma})
\end{itemize}
then the topology forces the f\/ixed points with $\Im\sigma>0$ to be saddle points and the only
f\/low possible is compatible with the topology of Fig.~\ref{fig1}.  The precise form of the f\/low
in Fig.~\ref{fig1} can be deformed, but the f\/ixed points cannot be moved.  

\begin{figure}[t]
\centerline{\includegraphics[width=6cm]{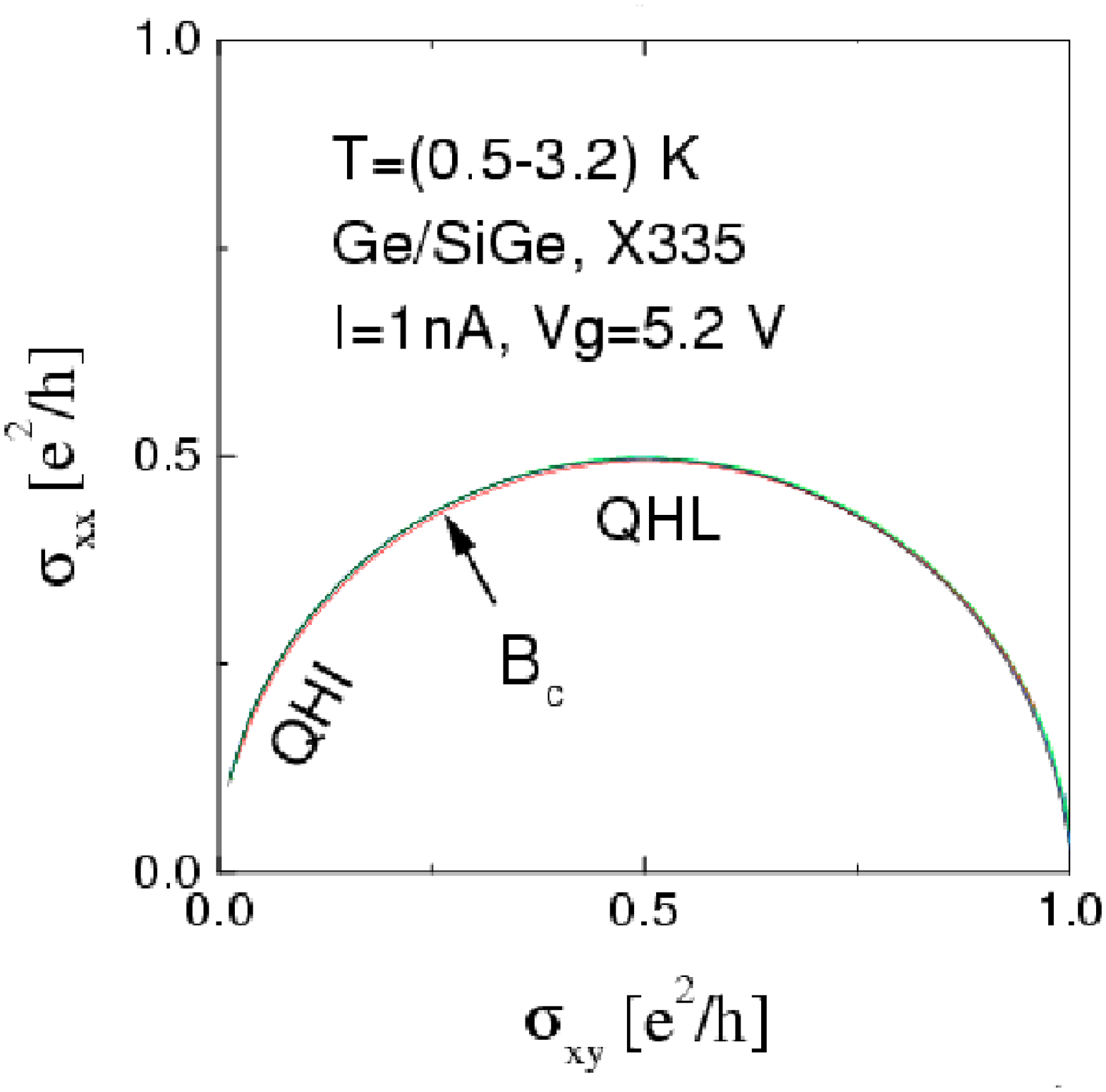}} \caption{\label{fig6}}
\end{figure}

An even stronger statement
can be made with one further assumption.  When a system is symmetric under particle-hole 
interchange one has symmetry under change of the sign of $\sigma_H$,
which is $\sigma\rightarrow 1-\bar\sigma$ for the complex conductivity.
This puts a reality condition on $\Sigma(\sigma,\bar\sigma)$ which, combined with the mathematical properties
of invariant functions of $\Gamma_0(2)$, can be used to show 
that the boundary of the fundamental domain in Fig.~\ref{fig5},
and its images under $\Gamma_0(2)$ must always be f\/low lines, \cite{semicircle}. Any deformation
of Fig.~\ref{fig1} must therefore leave invariant not only the f\/ixed points but also 
the vertical lines above integers on the
real axis and the semi-circles spanning rational numbers on the real line with odd denominators,
(as well as the images of these under $\Gamma_0(2)$)
 -- only the specif\/ic shape of the f\/low lines inside the fundamental domain, and their
images under $\Gamma_0(2)$, can be deformed.  In particular $\Gamma_0(2)$ symmetry provides
a remarkably robust derivation of the well-established experimental semi-circle law.
In many experiments the transition between two plateaux does follow a semi-circle
in the upper-half complex conductivity plane to a very high degree of accuracy,
Fig.~\ref{fig6} for example is taken from \cite{Hilkesemicircle}.  (Note however that the critical
point, as indicated by $B_c$ in Fig.~\ref{fig6}, is not at $\sigma=(1+i)/2$ as would be predicted
by $\Gamma_0(2)$ symmetry.  This will be discussed below.)
The scaling hypothesis of \cite{PruiskenScaling,SUnivWTPP} described above
suggests that, at low temperatures, $\sigma$
becomes a function of the scaling variable $\frac{\Delta B}{T^\kappa}$
and the f\/low is then forced onto one of these semi-circles as $T\rightarrow 0$. 
At a microscopic level the semi-circle law has been derived in one specif\/ic model \cite{Ruzin,RuzinFeng}, but modular
symmetry provides a very robust derivation, \cite{semicircle},
valid for any model which satisf\/ies the above
assumptions, and is therefore much more general than any specif\/ic model.
Any deviation from a semi-circular transition is an experimental signal that the sample
under investigation is not symmetric under particle-hole interchange. 
The general topology of the f\/low in Fig.~\ref{fig1}, at least for the integer QHE, was predicted
by Khmel'nitskii \cite{Khmelnitskii}, though the normalisation of the vertical axis was not
determined in that analysis and there was no extension to the fractional case\footnote{A similar
topology appears in the f\/low of one-dimensional clock models \cite{Asorey}, but there is no
obvious connection with the modular group in these models.}.

\begin{figure}[t]
\centerline{\includegraphics[width=5.5cm]{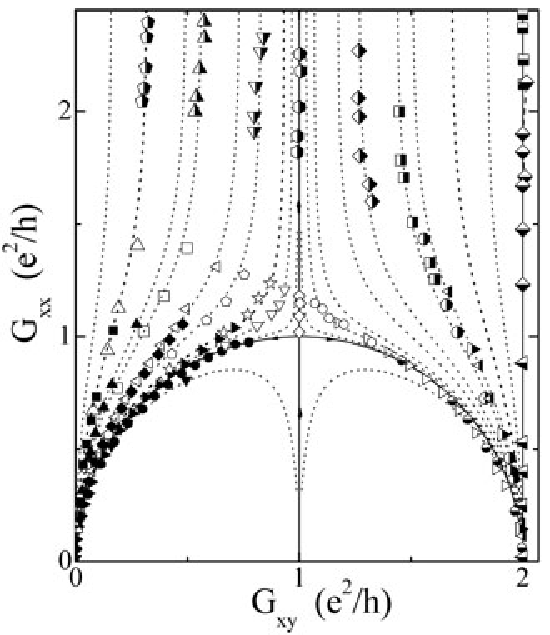}}
\vspace{-3mm}
 \caption{\label{fig7}}
\end{figure}

\begin{figure}[t]
\centerline{\includegraphics[width=10cm]{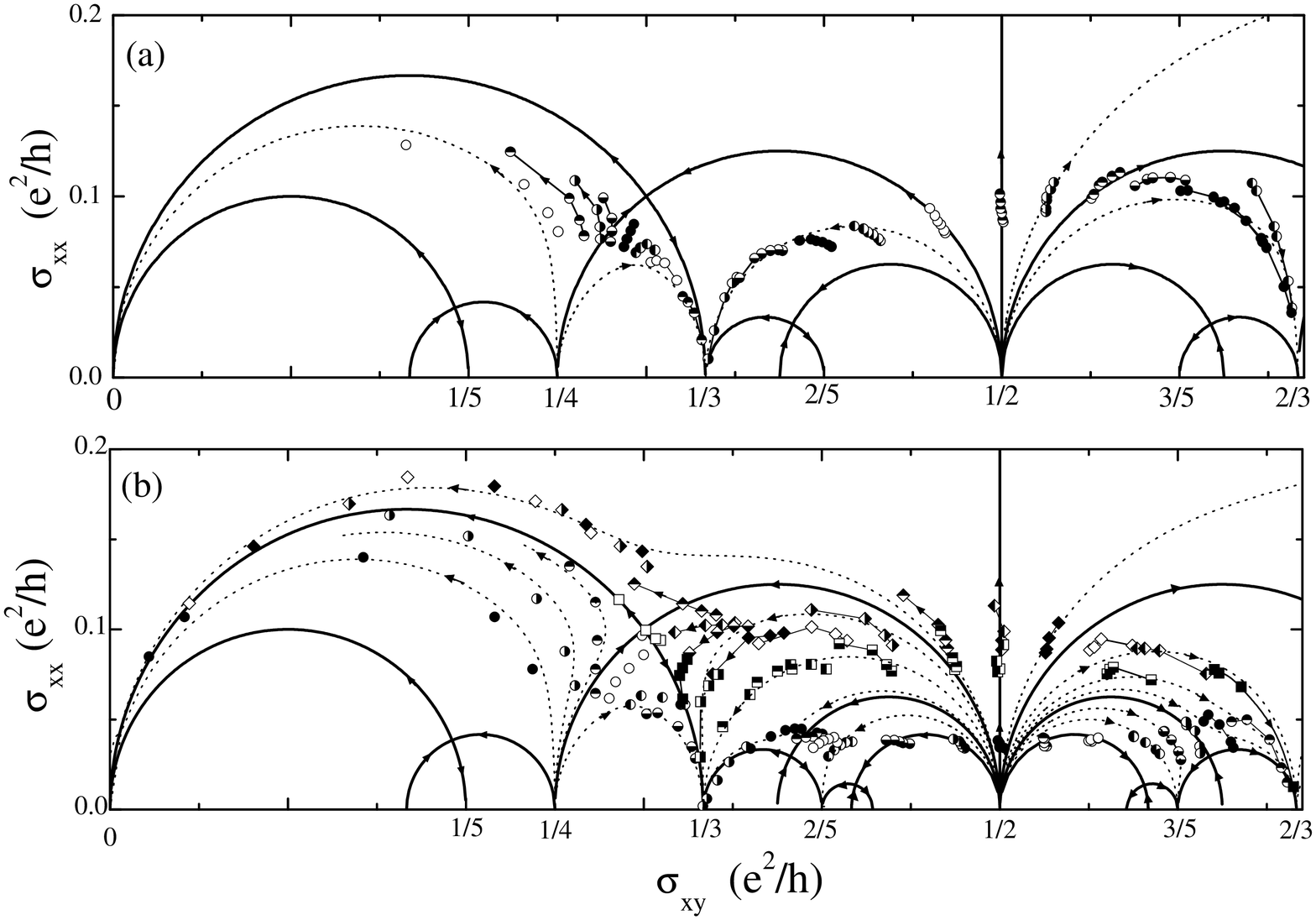}} 
\vspace{-3mm}
\caption{\label{fig8}}
\end{figure}

The specif\/ic f\/low for Fig.~\ref{fig1} was determined by supplementing the above assumptions with
one more condition, that the scaling function $\Sigma(\sigma)$ should be a meromorphic
function  in the argument $\sigma$\footnote{Strictly speaking it is meromorphic in the variable
$q=e^{i\pi\sigma}$ rather than in $\sigma$ itself.}.  This makes $\Sigma(\sigma)$ 
a modular form of weight $-2$
and the analytic form that approaches the stable f\/ixed points on the real axis most rapidly
is then exactly the same as that of ${\cal N}=2$ SUSY Yang--Mills in the previous section, 
up to an undetermined constant, namely
\begin{gather} \label{QHEflow}
\Sigma(\sigma)=\frac{2}{\pi i}\frac{1}{\bigl(\vartheta_3^4(\sigma)+\vartheta_4^4(\sigma)\bigr)},
\end{gather}
and this gives the f\/low plotted in Fig.~\ref{fig1}.
To date there is no physical reason for assuming that $\Sigma$ should be independent of $\bar\sigma$,
it is only motivated by mathematical analogy with SUSY Yang--Mills, but it does have the advantage
of giving an explicit form for the scaling function that can be visualised.  Any deformation away
from holomorphicity by including a non-meromorphic component is constrained by the considerations above
and cannot change the topology.  In fact the f\/low obtained using the meromorphic ansatz
does give remarkably good agreement with experiment and the
comparison is plotted in Figs.~\ref{fig7} and~\ref{fig8}, taken from \cite{MurzinZ,MurzinQ}
(in Fig.~\ref{fig7} the Landau levels are spin-degenerate and this has the ef\/fect of doubling the
conductivity -- see below and footnote~\ref{footnote14}).
Alternative, non-holomorphic forms, have recently been proposed \cite{LutkenRoss3,Lutken2}.

In summary, modular symmetry $\Gamma_0(2)$ applied to the quantum Hall ef\/fect for spin-split samples
leads to the following
predictions:
\begin{itemize}\itemsep=0pt
\item Universal critical points
are predicted at $\sigma_* = \frac12 (1+i)$ and its images under $\Gamma_0(2)$.
Critical exponents must be the same for
all f\/ixed points which are related by $\Gamma_0(2)$.
\item Exact f\/low lines in the $\sigma$ plane can be derived from $\Gamma_0(2)$
invariance plus invariance under particle-hole symmetry: $\sigma \to 
1 - \bar\sigma$. 
In particular the semi-circle law is a consequence of~$\Gamma_0(2)$ plus particle-hole symmetry.
\item Flow in the infrared is towards the real axis, terminating on the
real axis at attractive f\/ixed points at odd-denominator fractions. 
Even-denominator fractions form repulsive f\/ixed points
of the f\/low. 
\item The selection rule $|p_1 q_2 - p_2 q_1 | =1$ for allowed transition
between $\sigma=q_1/p_1$ and $\sigma=q_2/p_2$.
\end{itemize}

\begin{figure}[t]
\centerline{\includegraphics[width=8cm]{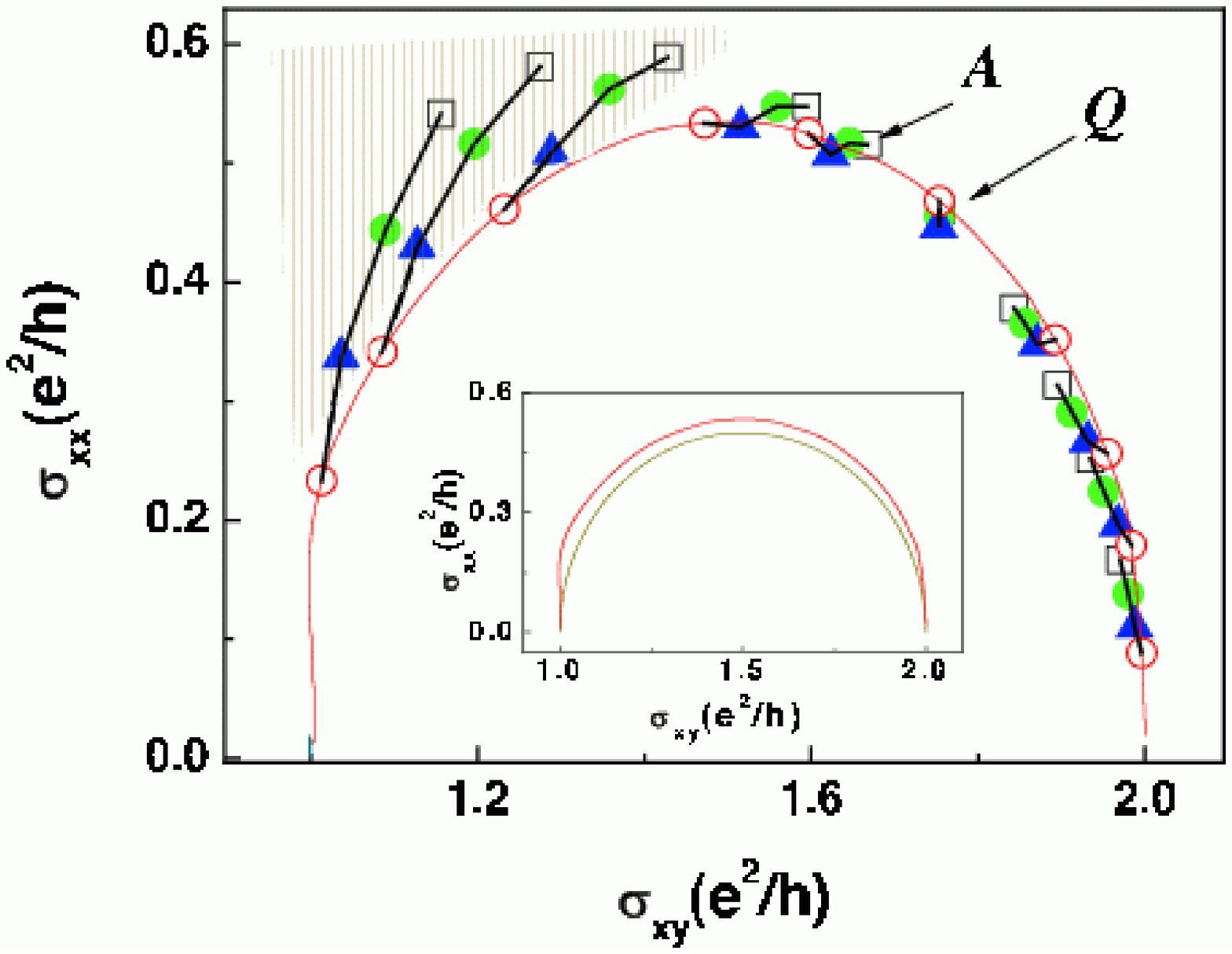}} 
\vspace{-2mm}
\caption{\label{fig9}}
\end{figure}

For QHE samples that are not
fully spin polarised one expects that the Landau levels can come in adjacent pairs and
${\cal T}$ in (\ref{Gamma02sigma}) should be replaced by ${\cal T}^2$, giving the generators of
$\Gamma(2)$ rather than $\Gamma_0(2)$, and this modif\/ies the experimental consequences.
$\Gamma(2)$ actually has no f\/ixed points above the real axis, so one cannot make any predictions
about the position of the second order phase transition between two plateaux, but if
there is particle-hole symmetry one still
has the semi-circle law and the f\/ixed point must lie somewhere on the semi-circle,~\cite{Gamma2}.
An experimental analysis of the relation between modular symmetry and Zeeman splitting 
is given in \cite{Huang1,Huang2,Huang3} and the results 
support the interpretation in~\cite{Gamma2}.
Fig.~\ref{fig9} is reproduced from~\cite{Huang3} and represents
a sample where the spin splitting is small, signif\/icantly less than the Landau level splitting 
but still non-negligible in the low Landau levels where the magnetic f\/ield is large, and
$\Gamma_0(2)$ symmetry is broken to $\Gamma(2)$, so the critical point, $Q$ in the f\/igure,
is not at the top of the semi-circle.  Nevertheless particle-hole symmetry is obeyed reasonably
well, as manifest by the close approximation to a semi-circle.  In the same sample, at lower
magnetic f\/ields and higher Landau levels the spin-splitting becomes negligible and the
sample becomes spin-degenerate.  In this situation $\Gamma_0(2)$ symmetry again applies, but
with the conductivity doubled because of the spin degeneracy in each Landau level.
The critical point in the $\nu=2 \rightarrow \nu=4$ transition is indeed at the top
of the semi-circle in the conductivity plane, as reported in~\cite{Huang3}\footnote{Because 
of the spin degeneracy the
transformations are $\sigma\rightarrow\sigma+2$ and $\sigma\rightarrow \frac{\sigma}{1-\sigma}$,
i.e.\ $\Gamma_0(2)$ acting on $\sigma/2$ is equivalent to $\Gamma^0(2)$ acting on $\sigma$,
just as for $N_f=3$ in SUSY.\label{footnote14}}.
It is possible that the explanation of the fact that $B_c$ is not at the
top of the semi-circle in Fig.~\ref{fig6} is due to the spins being poorly split, 
but there is another possible interpretation
for this $\nu: 1\rightarrow0$ transition~\cite{Gamma2}.  
This particular transition is a vertical line in the
complex $\rho$-plane above the point $\rho=1$ and a re-scaling of $\rho_{xx}$ can move the
critical point, which should be at $\rho_{xx}=1$ if $\Gamma_0(2)$ symmetry holds, to any value 
of $\rho_{xx}$ and hence to any point on the corresponding semi-circle in the $\sigma$-plane.
If the determination of the scale of $\rho_{xx}$
in \cite{Hilkesemicircle} has an error then Fig.~\ref{fig6} is still compatible with $\Gamma_0(2)$
symmetry. For the particular case of the $\nu:1\rightarrow0$ transition any re-scaling 
of $\rho_{xx}$ still gives a semi-circle in the $\sigma$-plane, but this is not true for other transitions.
The observation that the critical point is not at $\sigma=(3+i)/2$ for the $\nu:2\rightarrow1$
transition in \cite{Huang3}, as shown in Fig.~\ref{fig9}, cannot be explained by a re-scaling of the Ohmic
resistivity and therefore is interpreted as a breaking of $\Gamma_0(2)$ symmetry
down to $\Gamma(2)$ because the Landau levels, while not exactly degenerate, are not well 
split.
  
The group $\Gamma(2)$ was also analysed, from the point of view of its 
action on ground state wave-functions
rather than on complex conductivities, in \cite{Flohr} and on complex conductivities in 
\cite{Yves,Yves+,Yves2}.

$\Gamma_0(2)$ is the subgroup of the full modular group that is relevant for systems
with fermionic charge carriers in a strong perpendicular magnetic f\/ield with the spins well split.
Building on the work of \cite{KLZ}, and following a suggestion in \cite{Pryadko},
it was shown in \cite{Modular1}
that $\Gamma_0(2)$ should be the group relevant to 2-dimensional systems involving spin polarised fermionic
quasi-particles while Wilczek and Shapere's group $\Gamma_\theta$ should be relevant to 2-dimensional systems
involving bosonic quasi-particles. 
If the ef\/fective charge carriers are bosonic, e.g.\ 2-dimensional superconductors, one can 
obtain predictions from the fermionic case by using the f\/lux attachment transformation to add
an odd number of vortices to the quasi-particles.  For the case of a single unit of f\/lux this is
equivalent to conjugating $\Gamma_0(2)$ by ${\cal F}={\cal S}^{-1}{\cal T}{\cal S}$.
Since ${\cal F}^{-1} {\cal F}^2{\cal F}={\cal F}^2$ and ${\cal F}^{-1} {\cal T}{\cal F}={\cal F}^{-2}{\cal S}$ 
the resulting group is generated by $\cal S$ and ${\cal F}^2$, or equivalently $\cal S$ and ${\cal T}^2$, and is the
group $\Gamma_\theta$, \cite{WilczekShapere,Modular2,Modular1}.
Note that the ${\cal S}$ transformation in (\ref{GammaTheta}) requires $\vartheta=-s=\pm \eta$
with $\eta \rightarrow \infty$ in (\ref{KLZgamma}).  The f\/ixed point $\sigma=i$ of ${\cal S}$
is associated with a phase transition from an insulator to a superconductor~\cite{Fisher}
and a superconductor would have a double pole in its response function, associated
with spontaneous symmetry breaking and requiring a
mass term $m^2 A^2$ in the ef\/fective action.
A direct derivation of ${\cal S}$ transformation, in conjunction with particle-hole symmetry,
for a 2-dimensional superconductor was given in \cite{ReyZee}. 
$\Gamma_\theta$ consists of elements $\gamma$ of the form (\ref{abcd})
with either $a$, $d$ both odd and $b$, $c$ both even, or {\it vice versa}.
In this case the predictions are dif\/ferent from those of $\Gamma_0(2)$:
\begin{itemize}\itemsep=0pt
\item
Universal critical points are predicted for the f\/low at the f\/ixed points
for transitions between stable phases,
$\sigma_* = i$ and its images under $\Gamma_\theta$\footnote{This 
statement is for bosonic charge carriers with the
same electric charge as an electron -- 
it becomes $\sigma_* = i \tilde q^2$ and its images if the bosonic charge
carriers have charge $\tilde q$. In particular, the case $\tilde q=\pm 2$ applies if
the bosons are Cooper pairs, such as considered in~\cite{Fisher}.}. 

\item
The critical exponents at all f\/ixed points related by $\Gamma_\theta$
must all be the same. 

\item
Exact f\/low lines in the $\sigma$ plane are immediate consequences of 
$\Gamma_\theta$ invariance and particle-hole symmetry, \cite{Modular1}.
The results are again semi-circles or vertical lines
in the $\sigma$ plane, implying a semi-circle law for these
bosonic systems. 
\item
For nonzero magnetic f\/ields the f\/low as the temperature is reduced
is towards the real axis, terminating on the attractive 
f\/ixed points $\sigma = p/q$ with $pq$ even (as opposed to having
$q$ odd, as was the case for fermions). Fractions with odd $pq$ 
are repulsive f\/ixed points. 
\item
There is a selection rule that allowed 
fractions $p_2/q_2$ can be obtained from $p_1/q_1$ only if
$|p_1 q_2 - p_2 q_1|=1$.
\end{itemize}

\begin{figure}[t]
\centerline{\includegraphics[width=10.5cm]{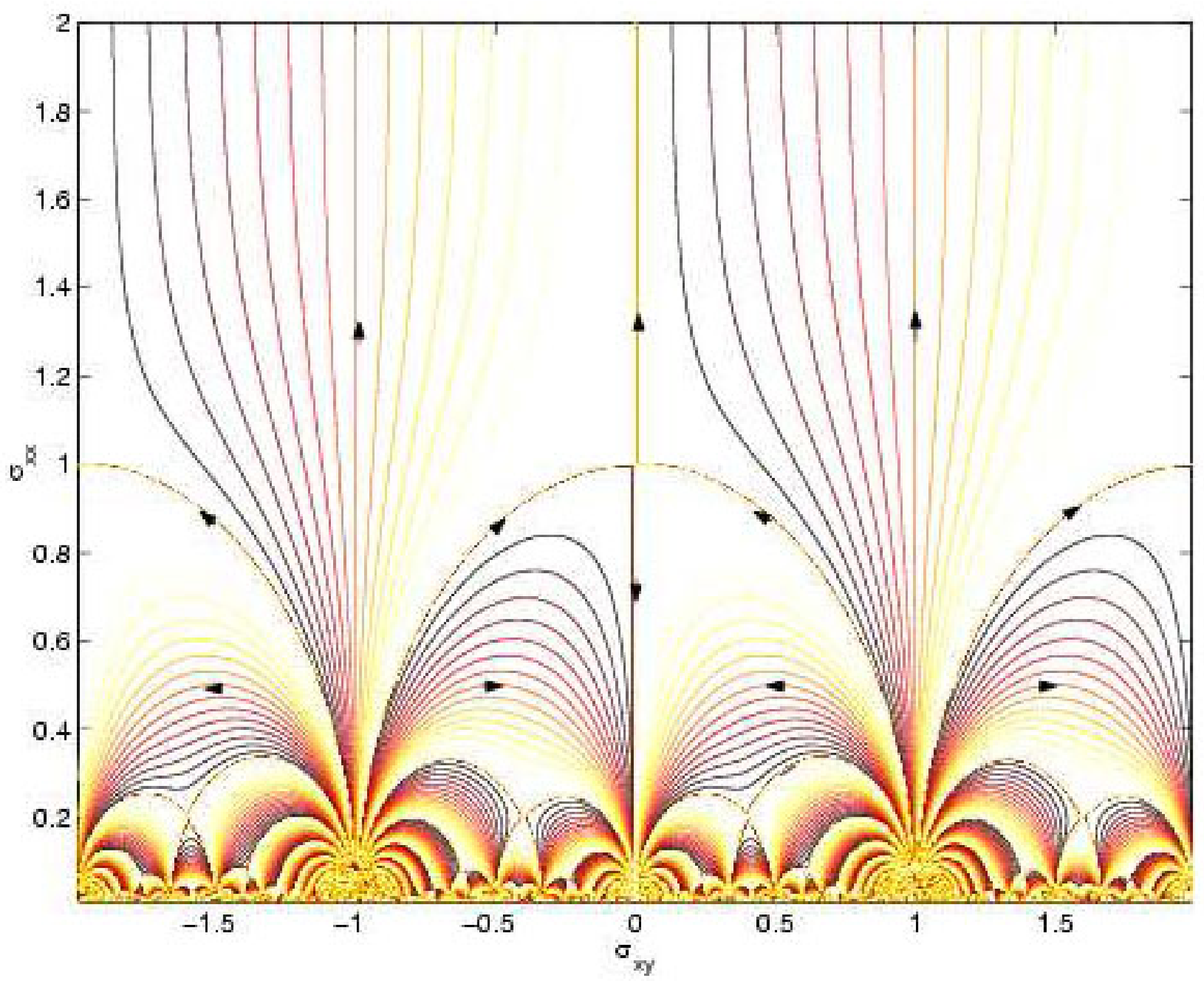}} 
\vspace{-2mm}
\caption{\label{fig10}}
\end{figure}

The resulting f\/low diagram for bosonic systems was f\/irst
given in \cite{Modular1} and is shown here in Fig.~\ref{fig10}.  It has a f\/ixed point
at $\sigma_*=i$, as predicted by Fisher \cite{Fisher}.
To date no 2-dimensional superconductors have been manufactured with a high enough mobility $\mu$
that ${\mu B}$ is close to unity for sustainable magnetic f\/ields,
but it is predicted in \cite{Modular1} that a
hierarchy with the above properties will be observed if such samples are ever manufactured. 

\section{Conclusions}

Modular symmetry is a generalisation of the Dirac quantisation condition for 
charge in QED.  Its mathematical foundation is strongest in supersymmetric systems, in particular
supersymmetric Yang--Mills, 
but a more realistic system, with a wealth of experimental data
to compare with, is the quantum Hall ef\/fect.  

Ef\/fective degrees of freedom are composite objects carrying topologically non-trivial gauge
f\/ield conf\/igurations (monopoles in SUSY Yang--Mills and vortices in the QHE).

For ${\cal N}=2$ SUSY $SU(2)$ Yang--Mills 
the subgroups of the modular group that are relevant for the scaling modular forms 
with dif\/ferent numbers of families are
\begin{gather*}
N_f=0, \qquad \Gamma^0(2), \nonumber \\ 
N_f=1, \qquad \Gamma(1),  \nonumber \\
N_f=2, \qquad \Gamma_0(2),  \\
N_f=3, \qquad \Gamma_0(4).\nonumber 
\end{gather*}
For the quantum Hall ef\/fect the relevant subgroups are
\begin{alignat*}{3}
&\hbox{spin-split samples}, \qquad && \Gamma_0(2), &  \\
&\hbox{spin-degenerate samples}, \qquad&& \Gamma^0(2), & \\ 
&\hbox{intermediate case}, \qquad&& \Gamma(2). &
\end{alignat*}

Modular symmetry has important experimental consequences for the
quantum Hall ef\/fect.  The Dirac--Schwinger--Zwanziger quantisation condition
manifests itself as a selection rule for transitions between allowed
quantum Hall plateaux,
\begin{gather*} |qp'-q'p|=1 
\end{gather*}
with Hall conductivities $q/p$ and $q'/p'$
in spin-split samples ($|qp'-q'p|=2$ for spin-degenerate samples).
Universal critical points in transitions between
plateaux for spin-degenerate and spin-split samples are an unavoidable consequence
of modular symmetry.
Together with particle-hole symmetry, when it is present, modular symmetry
immediately implies the semi-circle law for quantum Hall transitions -- 
any deviation from
the semi-circle law is most likely due to breaking of particle-hole symmetry in
the sample.  Fractional charges in the quantum Hall ef\/fect are related
to the Witten ef\/fect in 3+1-dimensional ${\cal N}=2$ supersymmetric Yang--Mills, where
massive excitations can carry electric charge $1/p$ with $p$ being the monopole number.

For bosonic pseudo-particles the group $\Gamma_\theta$ is
predicted to be the relevant group and this gives a suite of predictions similar
to the quantum Hall ef\/fect but dif\/fering in detail, as described in Section~4.

The connection between ${\cal N}=2$ SUSY and the quantum Hall 
ef\/fect that is exposed by modular symmetry is not yet understood, at the moment
it is merely at the level of an empirical observation 
that modular symmetry is relevant to scaling
in both these systems.  Constructing an explicit map relating one to the other
remains an open problem to date.
There is a known relation between ${\cal N}=4$ SUSY and the quantum Hall ef\/fect,
\cite{Berenstein,LLM,SheikhJabbari}, which may prove to be useful starting point,
perhaps via SUSY breaking,
but since the perturbative $\beta$-function for ${\cal N}=4$ SUSY vanishes there is
no clear connection with scaling at present.
There are however certain features which both ${\cal N}=2$ SUSY and the QHE have in common.
They all have only two relevant couplings: one, $y$, associated with the dynamical kinetic 
term, which must be positive for stability reasons; and one, $x$, associated with
a topological term.  These are combined into the complex parameter, $\tau=x+iy$, on the
upper half-complex plane on which the modular group, or a subgroup thereof, acts.

An important feature that these systems have in common is the existence of topologically
non-trivial f\/ield conf\/igurations, monopoles in SUSY Yang--Mills and vortices in the QHE,
which can bind to the charge carriers to form pseudo-particles carrying
magnetic charge: dyons in SUSY Yang--Mills and in the QHE fermions which can be viewed as
bosons with an odd number of f\/lux units attached.
The rational nature of the attractive f\/ixed points on the real axis $q/p$ is then related
to the magnetic charge $p$.  
The weak coupling regime, where perturbation theory might be expected to be useful, can be mapped
by modular symmetry to the strong coupling regime, where much interesting physics lies.

It seems that modular symmetry is not just an accident of one type of system, or even
a family of systems such as supersymmetric f\/ield theories, but in fact is a more
general phenomenon and it may yet prove even more powerful in understanding
the physics of strongly interacting systems.

\subsection*{Acknowledgements} It is a pleasure to thank my long-term collaborator Clif\/f Burgess
for many useful discussions on the quantum Hall ef\/fect. 
This work was partly supported by Enterprise Ireland Basic Research grant SC/2003/415.

\pdfbookmark[1]{References}{ref}
\LastPageEnding

\end{document}